\documentclass[]{emulateapj}
\usepackage{lscape}
\usepackage{subfigure}

\newcommand{\noprint}[1]{}

\begin{document}

\title{Characterizing the Cool KOI\lowercase{s} III.  KOI-961: A Small Star with Large Proper Motion and Three Small Planets}

\author{Philip S. Muirhead,\altaffilmark{1,2} 
John Asher Johnson,\altaffilmark{1,14} 
Kevin Apps,\altaffilmark{3} 
Joshua A. Carter,\altaffilmark{4,5}
Timothy D. Morton,\altaffilmark{1}
Daniel C. Fabrycky\altaffilmark{6,5},
J. Sebastian Pineda,\altaffilmark{1} 
Michael Bottom,\altaffilmark{1} 
B{\'a}rbara Rojas-Ayala,\altaffilmark{7}
Everett Schlawin,\altaffilmark{8}
Katherine Hamren,\altaffilmark{6}
Kevin R. Covey,\altaffilmark{8,5}
Justin R. Crepp,\altaffilmark{1}
Keivan G. Stassun,\altaffilmark{9, 10, 11}
Joshua Pepper,\altaffilmark{9}
Leslie Hebb,\altaffilmark{9}
Evan N. Kirby,\altaffilmark{1,5}
Andrew W. Howard,\altaffilmark{12}
Howard T. Isaacson,\altaffilmark{12}
Geoffrey W. Marcy,\altaffilmark{12}
David Levitan,\altaffilmark{1}
Tanio Diaz-Santos,\altaffilmark{13}
Lee Armus,\altaffilmark{13}
James P. Lloyd\altaffilmark{8}}

\altaffiltext{1}{Department of Astronomy, California Institute of Technology, 1200 East California Boulevard, MC 249-17, Pasadena, CA  91125, USA}

\altaffiltext{2}{Email: {\tt philm@astro.caltech.edu}}

\altaffiltext{3}{75B Cheyne Walk, Surrey, RH6, 7LR, United Kingdom}
\altaffiltext{4}{Harvard-Smithsonian Center for Astrophysics, 60 Garden Street, Cambridge, MA 02138, USA}
\altaffiltext{5}{Hubble Fellow}
\altaffiltext{6}{UCO/Lick, University of California, Santa Cruz CA 95064}
\altaffiltext{7}{Department of Astrophysics, American Museum of Natural History, Central Park West at 79th Street, New York, NY 10024, USA.}
\altaffiltext{8}{Department of Astronomy, Cornell University, 122 Sciences Drive, Ithaca, NY 14583, USA}
\altaffiltext{9}{Department of Physics and Astronomy, Vanderbilt University, Nashville, TN 37235, USA}
\altaffiltext{10}{Fisk University, Department of Physics, 1000 17th Ave. N., Nashville, TN 37208}
\altaffiltext{11}{Massachusetts Institute of Technology, Department of Physics, 77 Massachusetts Ave., Cambridge, MA  02139}

\altaffiltext{12}{Astronomy Department, University of California, Berkeley, CA 94720, USA}
\altaffiltext{13}{Spitzer Science Center, California Institute of Technology, 1200 East California Boulevard, Pasadena, CA  91125, USA}
\altaffiltext{14}{NASA Exoplanet Science Institute (NExScI), CIT Mail Code 100-22, 770 South Wilson Avenue, Pasadena, CA 91125}

\date{}

\begin{abstract}

We present the characterization of the star KOI 961, an M dwarf with transit signals indicative of three short-period exoplanets, originally discovered by the {\it Kepler} Mission.  We proceed by comparing KOI 961 to BarnardÕs Star, a nearby, well-characterized mid-M dwarf.  By comparing colors, optical and near-infrared spectra, we find remarkable agreement between the two, implying similar effective temperatures and metallicities.  Both are metal-poor compared to the Solar neighborhood, have low projected rotational velocity, high absolute radial velocity, large proper motion and no quiescent H$\alpha$ emission--all of which is consistent with being old M dwarfs.  We combine empirical measurements of Barnard's Star and expectations from evolutionary isochrones to estimate KOI 961's mass (0.13 $\pm$ 0.05 $\rm M_\Sun$), radius (0.17 $\pm$ 0.04 $\rm R_\sun$) and luminosity (2.40  $\times$ $10^{-3.0 \pm 0.3 }$ $\rm L_\sun$).  We calculate KOI 961's distance (38.7 $\pm$ 6.3 pc) and space motions, which, like Barnard's Star, are consistent with a high scale-height population in the Milky Way.  We perform an independent multi-transit fit to the public {\it Kepler} light curve and significantly revise the transit parameters for the three planets.  We calculate the false-positive probability for each planet-candidate, and find a less than 1 \% chance that any one of the transiting signals is due to a background or hierarchical eclipsing binary, validating the planetary nature of the transits.  The best-fitting radii for all three planets are less than 1 $R_\Earth$, with KOI 961.03 being Mars-sized ($R_P$ = 0.57 $\pm$ 0.18 $\rm R_\Earth$), and they represent some of the smallest exoplanets detected to date.

\end{abstract}


\maketitle

\section{Introduction}

In February of 2011 the {\it Kepler} Mission announced 997 stars that show light curves consistent with the presence of transiting planets \citep[][]{Borucki2011}.  Many of the {\it Kepler} Objects of Interest, or KOIs, host planet-candidates that are terrestrial-sized, and represent some of the smallest exoplanets detected around main-sequence stars if they are confirmed.  \citet{Morton2011} found that many of the planets have low {\it a priori} probabilities of being false positives, and careful follow-up studies will validate the planetary nature of many of the transit signals.

KOI 961 ({\it Kepler} ID: 8561063) is particularly interesting: an M dwarf star with large proper motion \citep[$\mu$ = 431 $\pm$ 8 {\it mas/yr},][]{Lepine2005} and transit signals indicative of three exoplanets, all with orbital periods of less than 2 days.  The inferred radii of the planet-candidates depends directly on the assumed radius of the host star.  Unfortunately, the stellar parameters of KOI 961 are highly uncertain.  \citet{Borucki2011} listed an effective temperature of 4188 K and a radius of 0.68 $\rm R_\sun$ based on photometric analysis from the {\it Kepler} Input Catalog \citep[KIC, ][]{Batalha2010}.  However, \citet{Brown2011} warned that the stellar radii listed in the KIC have uncertainties of 30 to 50\%, and that the radii for M dwarfs are particularly untrustworthy.  Asteroseismology studies of {\it Kepler} objects by \citet{Verner2011} also indicate biases in the surface gravity and radius estimates in the KIC.   

Typically, M dwarf masses and radii are estimated by combining stellar luminosity with reliable mass-luminosity relations \citep[e.g.][]{Delfosse2000} and mass-radius relations predicted by stellar evolutionary models \citep[e.g.][]{Baraffe1998}, often with corrections to account for discrepancies between measured and predicted radii \citep[e.g.][]{Torres2007}.  Unfortunately, KOI 961 does not have a parallax measurement in the literature, which is necessary to estimate its luminosity and hence mass.

In a previous paper \citep[][hereafter Paper 1]{Muirhead2011b}, we acquired a $K$-band spectrum of KOI 961 as part of a survey of low-mass KOIs.  The spectrum confirmed that it is a dwarf and not a giant based on the shallowness of the CO absorption band, and we reported an effective temperature of 3200 $\pm$ 50 K and an overall metallicity ([M/H]) of -0.33 $\pm$ 0.12 using the spectroscopic indices and calibrations of \citet[][hereafter RA11]{Rojas2011}.  KOI 961 was the coolest M dwarf in our survey, and it is likely the least massive KOI in the February 2011 {\it Kepler} data release.  We interpolated the effective temperature and metallicity of KOI 961 onto the 5-Gyr Solar-metallicity evolutionary isochrones of \citet{Baraffe1998}, providing a stellar radius estimate of 0.19 $\rm R_\sun$ and a mass estimate of 0.16 $\rm M_\sun$.  The stellar parameters derived for the low-mass KOIs in Paper 1 are spectroscopic, and are more accurate than the KIC values, but depend strongly on the accuracy of evolutionary isochrones for low-mass stars.

In this paper we present measurements of the stellar parameters of KOI 961 by direct comparison to a well-studied low-mass star: Barnard's Star, or Gl 699.  Barnard's Star is the second nearest main-sequence star system to the Sun at a distance of 1.824 $\pm$ 0.005 pc \citep{vanleeuwen2007}.  Discovered by \citet{Barnard1916}, his ``Small Star with Large Proper Motion'' is classified as an M4 dwarf \citep[][]{Reid1995}, and \citet{Leggett1992} showed that Barnard's Star's kinematic and photometric properties are consistent with an ``old disk-halo'' star.  It has low projected rotational velocity \citep[$V\sin(i)<$ 2.5 km $\rm s^{-1}$,][]{Browning2010}, infrequent H$\alpha$ flaring with no H$\alpha$ emission during quiescence \citep[e.g.][]{Paulson2006} and lies below the average main sequence \citep{Giampapa1996}; all of which are consistent with an old, metal-poor M dwarf.  Being nearby, BarnardÕs Star is among the most studied and best characterized M dwarfs despite its extreme dissimilarity to the Sun.  It has an empirically-measured stellar radius using optical long-baseline interferometry \citep{Lane2001}, a carefully measured bolometric luminosity \citep{Dawson2004}, and an accurate {\it Hipparcos} parallax, which, when combined with photometry and the empirically-derived mass-luminosity relations of \citet{Delfosse2000}, provides an accurate measurement of its mass.

RA11 report metallicities for Barnard's Star of [M/H] = -0.27 $\pm$ 0.12 and [Fe/H] = -0.39 $\pm$ 0.17, and an effective temperature of 3266 $\pm$ 29 K.  The same methods were used to measure similar values of KOI 961.  This motivates a side-by-side comparison of the two objects, for the purpose of measuring the stellar parameters of KOI 961.  In Section \ref{observables}, we perform a qualitative comparison of the two stars' spectra, finding them to be very similar. We also compare various spectroscopic indices and argue that KOI-961 is slightly metal-poor compared to Barnard's Star, yet shares its effective temperature, consistent with the $K$-band measurements.  

In Section \ref{parameters} we perform a more quantitative assessment of KOI-961's stellar properties relative to Barnard's Star. We start with empirical measurements of the mass, radius and luminosity of Barnard's Star and appeal to stellar evolution models to estimate the relative difference between the two stars' physical parameters. In this manner we find that KOI-961 is slightly smaller and less massive than Barnard's Star. We also estimate KOI-961's distance and space motion and show they are consistent with a high scale-height, or ``thick-disk,'' population in the Milky Way.  

In Section \ref{transit} we refit transit curves for the three planets using the public {\it Kepler} data, revise the transit parameters, and revise the planet parameters, finding that all three are sub-Earth-sized.  In Section \ref{planets} we calculate the false-positive probability for the transit signals based on {\it a priori} statistics of eclipsing binaries and seeing-limited images, historical images, and the high-resolution optical spectrum of KOI 961, and we validate the planetary-nature of the transits.  In Section \ref{dynamics} we discuss the masses and dynamical configuration of the three planets.  In Section \ref{discussion} we discuss the implications given that KOI 961 is metal-poor, and one of only a few mid-M stars surveyed by the {\it Kepler} Mission.

\section{Comparison of Observables}\label{observables}

\subsection{Photometry}\label{photometry}

Digital photometric measurements of KOI 961 include $griz$ photometry from the KIC \citep[][]{Batalha2010} and 2MASS $JHK$ photometry \citep[][]{Cutri2003}.  Digital photometric measurements of Barnard's Star include $UVBRI$ photometry by  \citet{Koen2010} and 2MASS $JHK$ photometry.  The 2MASS $J$- and $H$-band magnitudes of Barnard's Star are near the saturation limit \citep{Skrutskie2006}, and produce colors that are inconsistent with the spectral type from \citet{Reid1995}.  With this in mind, we used the $J_{\rm CIT}$- and $H_{\rm CIT}$-band measurements of Barnard's Star by \citet{Leggett1992}, which we converted to 2MASS $J$ and $H$ magnitudes using the transformations of \citet{Carpenter2001}.

We obtained Johnson $B$- and $V$-band and Cousins $Rc$-band measurements of KOI 961 with the Hereford Arizona Observatory (HAO) 11-inch telescope on UT 2011 November 17, using standard star fields from \citet{Landolt1992} for calibration.  Three \citet{Landolt1992} star fields were used containing a total of 45 total stars with accurate $B$- and $V$-band photometry, and 40 stars with accurate $Rc$-band photometry.  We acquired 108 $B$- and $V$-band measurements of reference stars, and 94 $R_C$-band measurements to calibrate the measurements of KOI 961.  \citet{Gary2007} describes the calibration procedures for HAO photometric measurements.  The measurements and uncertainties are listed in Table \ref{hao}, along with available 2MASS $J$, $H$ and $K_{\rm S}$ measurements.

\begin{table}
\begin{center}
\caption{Photometry of KOI 961\label{hao}} 
\begin{tabular}{l c c c} 
\hline\hline                        
Band & Magnitude & System \\ [0.5ex] 
\hline                  
  $B$  & 17.96  $\pm$ 0.14 & Johnson\footnotemark[1] \\
    $V$  & 16.124 $\pm$ 0.055 & Johnson\footnotemark[1]\\
    $R_C$ & 15.055 $\pm$ 0.051 & Cousins\footnotemark[1]\\
  $J$  & 12.177  $\pm$ 0.021 & 2MASS\footnotemark[2]\\
    $H$  & 11.685 $\pm$ 0.018 & 2MASS\footnotemark[2]\\
    $K_{\rm S}$ & 11.465 $\pm$ 0.018 & 2MASS\footnotemark[2]\\

    \hline 
\footnotetext[1]{Obtained with the HAO 11-inch telescope.}
\footnotetext[2]{\citet{Cutri2003,Skrutskie2006}}
\end{tabular}
\end{center}
\end{table}

\begin{figure}
\centering
\includegraphics[width=3.3in]{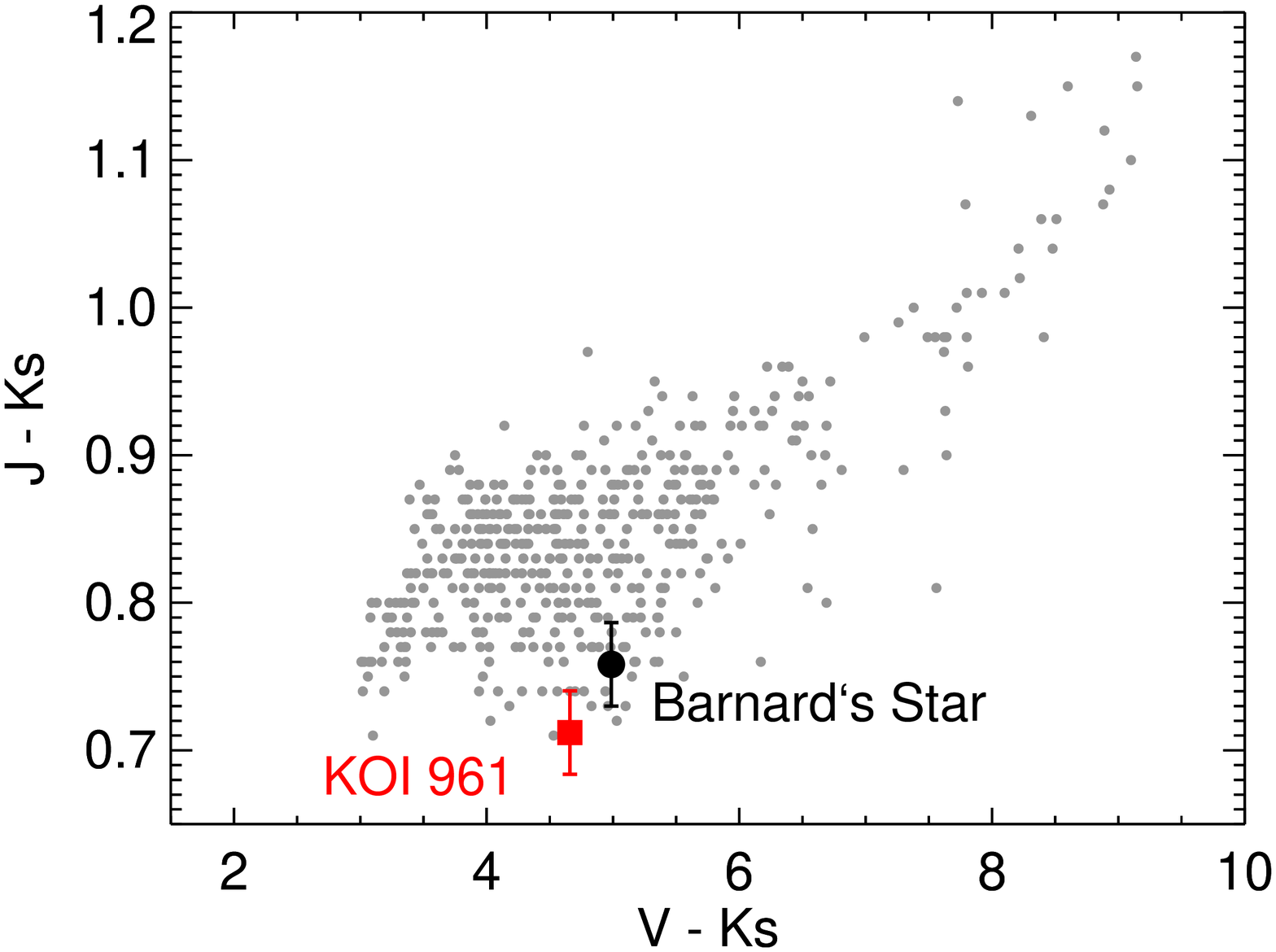}
\caption{Color-color plot showing $J$-$K_{\rm S}$ vs $V$-$K_{\rm S}$ for stars within 20 pc with $V-K \ge $ 3.0 (spectral types later than K7), trigonometric parallax uncertainty of $\le 5 \%$ \citep{vanleeuwen2007, vanaltena2001}, $V$ error of $\le$ 0.1 \citep{Koen2010} and 2MASS $J$ and $K_S$ uncertainties of $\le$ 0.05 \citep{Cutri2003, Skrutskie2006}.  Note the difference in scales for the two axes.  Included are Barnard's Star ({\it black circle}) and KOI 961 ({\it red square}), which are both at the lower edge of the distribution, owing to their low metallicity.  The similar  $V$-$K_{\rm S}$ and $J$-$K_{\rm S}$ colors for the two stars indicate similar effective temperature and metallicity \citep[e.g.][]{Leggett1992,Johnson2009}, consistent with the $K$-band measurements in Paper 1 and by RA11.\label{j_k}}

\end{figure}

\begin{figure}

\hspace{0.5cm}
\centering
\includegraphics[width=3.3in]{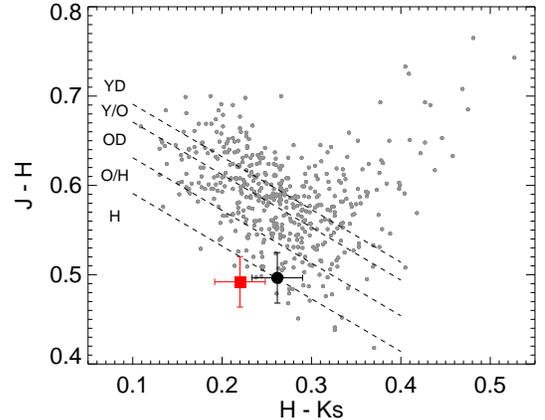}
\caption{Color-color plot showing $J$-$H$ vs. $H$-$K_{\rm S}$ for nearby M dwarfs, Barnard's Star ({\it black circle}) and KOI 961 ({\it red square}).  As in Figure \ref{j_k} we use $J$ and $H$ values from \citet{Leggett1992} for Barnard's Star out of concern for saturation effects.  We include the population designations of \citet{Leggett1992}, which relate to the spread in metallicity: young disk (YD), young-old disk (Y/O), old disk (OD), old disk-halo (O/H) and halo (H).  Both KOI 961 and Barnard's Star are in old and metal-poor populations.\label{h_k}
}
\end{figure}

Figures \ref{j_k} and \ref{h_k} show $J$-$K_{\rm S}$ vs $V$-$K_{\rm S}$ and $J$-$H$ vs. $H$-$K_{\rm S}$ for a sample of nearby M dwarfs, with Barnard's Star and KOI 961 indicated.  The $V$-$K_{\rm S}$ color is sensitive to both effective temperature and metallicity \citep{Leggett1992,Johnson2009}.  In \citet[][hereafter Paper 2]{Johnson2011} we identified the metallicity sensitivity of $J$-$K_{\rm S}$.  The $J$-$H$ vs. $H$-$K_{\rm S}$ color-color regime has also been shown to distinguish metallicities and ages of M dwarfs, and we include the population designations of \citet{Leggett1992} in Figure \ref{h_k}: young disk (YD), young-old disk (Y/O), old disk (OD), old disk-halo (O/H) and halo (H).

Barnard's Star and KOI 961 show remarkable agreement in both color-color plots.  This indicates that KOI 961 and Barnard's Star have similar effective temperatures and metallicities, and are metal-poor compared to the Solar neighborhood, consistent with the $K$-band spectroscopic measurements in Paper 1 and by RA11.  KOI 961 has a slightly bluer $V$-$Ks$ color (0.33 $\pm$  0.04), which can be due to either higher effective temperature or lower metallicity.  Medium-resolution optical spectroscopy of the the stars indicates the latter, and we discuss this in the following section.

\subsection{Medium-Resolution Optical Spectroscopy}

We observed KOI 961 on UT 22 July 2011 with the Double Spectrograph on the Palomar Observatory 200-inch Hale Telescope: a two-channel (Red/Blue), low-to-medium resolution spectrograph \citep{Oke1982}.  We used settings that provided spectral coverage from 5300 $\rm \AA$ to 10000 $\rm \AA$ in the red channel, with 10 $\rm \AA$-wide resolution elements.  At the time, the efficiency of the red channel limited the usable spectral coverage to 5300 to 8000 $\rm \AA$.  The blue channel exposure containes significantly less signal, so we do not include those data.  We observed the white dwarf standard BD +284211 at the same airmass to calibrate the throughput and telluric absorption in the raw spectrum.  We used a theoretical  spectrum of BD +284211 from \citet{Stone1977} for the calibration.  We used available dome lamps for flat-fielding and wavelength calibration, and reduced the data using the NOAO {\tt doslit} package, written in IRAF.  The final spectrum has a signal-to-noise ratio of roughly 350 at 6500 $\rm \AA$.

\begin{figure}[]
\begin{center}
\includegraphics[width=3.3in]{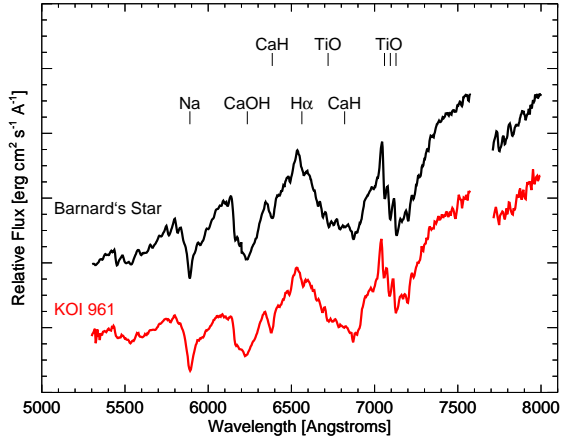}
\caption{Comparison of moderate-resolution optical spectra of KOI 961 and Barnard's Star.  {\it Red}: Double Spectrograph spectrum of KOI 961.  {\it Black}: Spectrum of Barnard's Star from \citet{Dawson2004}.  The spectra were renormalized, and the spectrum of Barnard's Star is offset for comparison.  The stars have remarkably similar spectra indicating similar spectral type and metallicity.  To quantitatively compare the stars we measured the PMSU spectroscopic indices (see Table \ref{indices} and Figure \ref{indices_plot}).  Neither spectrum shows H$\alpha$ emission, but Barnard's Star is known to have infrequent H$\alpha$ flaring \citep{Paulson2006}.}
\label{dbspec}
\end{center}
\end{figure}

\citet[][]{Dawson2004} combined spectra of Barnard's Star taken over many wavelength regimes in order to accurately measured the star's bolometric luminosity.  We show a region of their combined spectrum and our spectrum of KOI 961 in Figure \ref{dbspec}.  The spectra are strikingly similar, aided by the fact that they have similar resolution.  To quantitatively compare them, we calculated the spectroscopic indices of the Palomar-Michigan State University Nearby-Star Spectroscopic Survey \citep[PMSU, e.g.][]{Reid1995, Hawley1997, Gizis2002}.  We used the band definitions and index formula introduced by \citet{Reid1995}, who also report the indices for Barnard's Star.

Table \ref{indices} lists the PSMU spectroscopic indices for Barnard's Star from \citet{Reid1995}, our calculations using the \citet[][]{Dawson2004} spectrum of Barnard's Star, and our calculations using our spectrum of KOI 961.  The difference between our calculated indices for Barnard's Star and the \citet{Reid1995} values is likely due to the spectra having different resolution.  The standard deviation in the difference between our calculated indices of Barnard's Star and the PMSU indices is 0.05, which we used to estimate the uncertainty in the indices.  

\begin{figure}[]
\begin{center}
\includegraphics[width=3.3in]{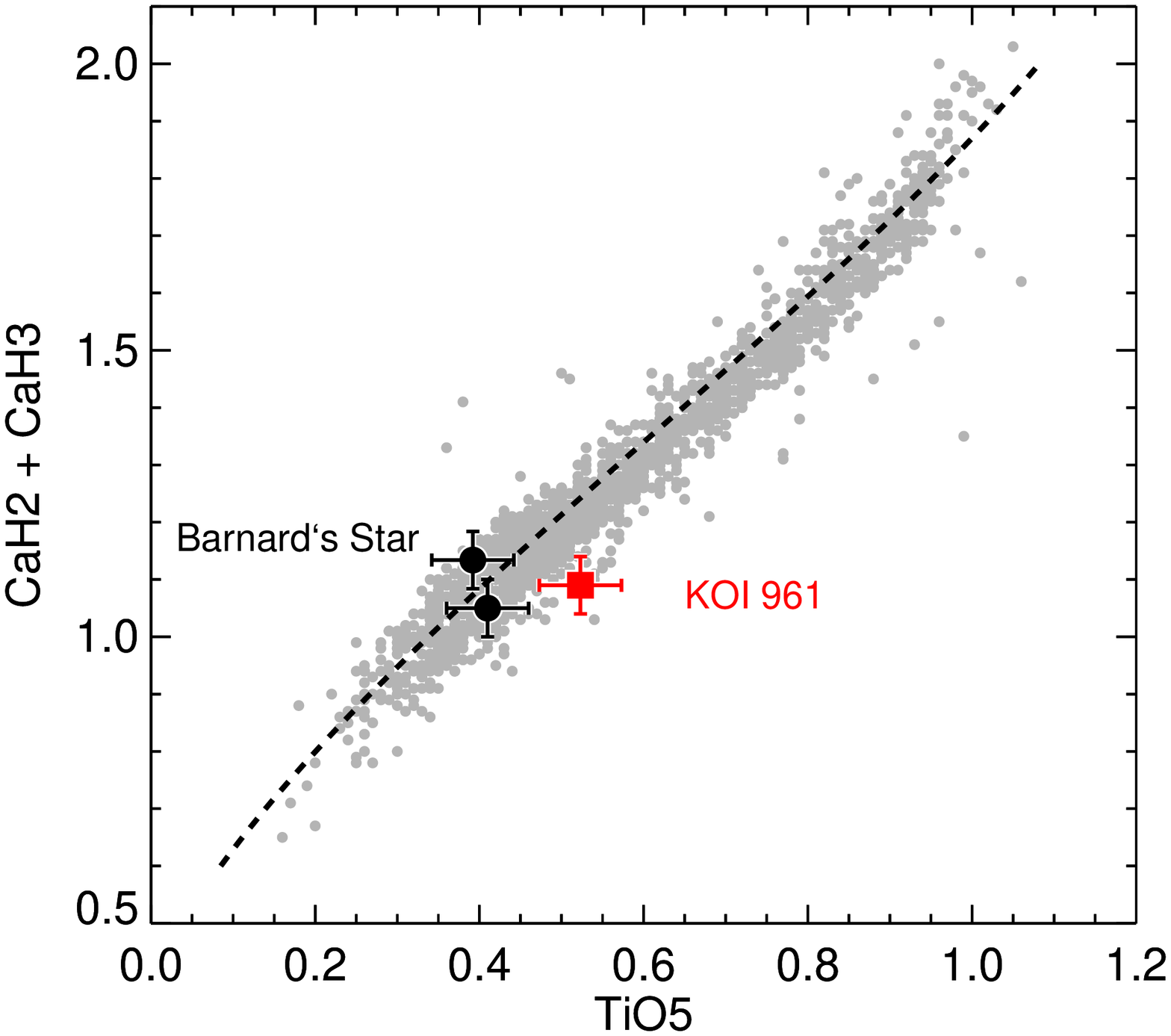}
\caption{PMSU spectral indices CaH2 + CaH3 vs. TiO5 [see Figures 1, 3, 6 and 8 of \citet{Lepine2007}], including all stars with values reported in the PMSU survey \citep{Reid1995,Hawley1997}, two estimates for Barnard's Star ({\it two black circles}, see text), KOI 961 ({\it red square}), and the mean fit to the Solar-metallicity disk dwarfs from \citet[][{\it dashed line}]{Lepine2007}.  KOI 961 has matching CaH depths but shallower TiO depths compared to Barnard's Star.  Using the interpretation \citet[][]{Lepine2007}, this indicates that KOI 961 has similar effective temperature but slightly lower metallicity compared to Barnard's Star, consistent with the $K$-band measurements from RA11 and Paper 1.}
\label{indices_plot}
\end{center}
\end{figure}

KOI 961 has very similar CaH indices to Barnard's Star, but systematically higher TiO indices, indicating {\it weaker} TiO absorption.  TiO absorption has long been the standard for spectral typing of M dwarfs, and is sensitive to stellar effective temperature \citep[e.g.][]{Reid1995}.  However, \citet{Lepine2007} showed that TiO is sensitive to both temperature {\it and} metallicity, and showed that CaH is a better diagnostic for effective temperature and spectral type.  The relationship between TiO and CaH for nearby stars, Barnard's Star and KOI 961 is illustrated in Figure \ref{indices_plot}, showing CaH2 + CaH3 vs. TiO5 [see Figures 1, 3, 6 and 8 of \citet{Lepine2007}].  KOI 961 and Barnard's Star have similar CaH2 + CaH3 values within their uncertainties, indicating similar effective temperature.  The position of KOI 961 to the right of Barnard's Star implies lower metallicity, consistent with the slightly bluer $V$-$K$ color.

The \citet{Lepine2007} TiO and CaH relations are diagnostics for estimating the relative differences in effective temperature and metallicity, and useful for distinguishing dwarfs from subdwarfs.  They were not calibrated to empirical metallicity or effective temperate measurements.  For {\it calibrated} measurements, we used the $K$-band spectroscopic indices of RA11.

\begin{table}
\begin{center}
\caption{Comparison of PMSU Spectroscopic Indices\label{indices}.} 
\begin{tabular}{l c c c} 
\hline\hline                        
Index & \multicolumn{2}{c}{--- Barnard's Star ---} & KOI 961 \\ [0.5ex] 
 & 1 & 2 & \\
\hline                  
	 
TiO 1 &  0.75 & 0.85 &  0.85\\
TiO 2 &  0.64 & 0.63 &  0.75\\
TiO 3 &  0.68 &  0.73 &  0.81\\
TiO 4 &  0.64 & 0.62 &  0.71\\
TiO 5 &  0.41 & 0.39 &  0.52\\
CaH 1 &  0.74 & 0.84 &  0.77\\
CaH 2 &   0.39 & 0.45 &  0.42\\
CaH 3 &  0.66  & 0.68 &  0.67\\
CaOH &  0.37 & 0.45 &  0.54\\
$\zeta_{\rm TiO/CaH}$\footnotemark[3] & 1.37 & 1.87 & 1.26\\
$\zeta_{\rm TiO/CaH}$\footnotemark[4] & 1.43 & 2.02 & 1.33\\
\hline 
\end{tabular}
\footnotetext[1]{Values from \citet{Reid1995}.}
\footnotetext[2]{Values calculated in this work using the spectrum from \citet{Dawson2004}.  Comparing the values to \citet{Reid1995}, we estimate 1-$\sigma$ uncertainties of 0.05 for all indices.}
\footnotetext[3]{The $\zeta_{\rm TiO/CaH}$ index is introduced in \citet{Lepine2007}.}
\footnotetext[4]{We also report the $\zeta_{\rm TiO/CaH}$ index using the coefficients from \citet{Dhital2011}.}
\end{center}
\end{table}

\subsection{Medium-Resolution Near-Infrared Spectroscopy}

\begin{figure}[]
\begin{center}
\includegraphics[width=2.8in]{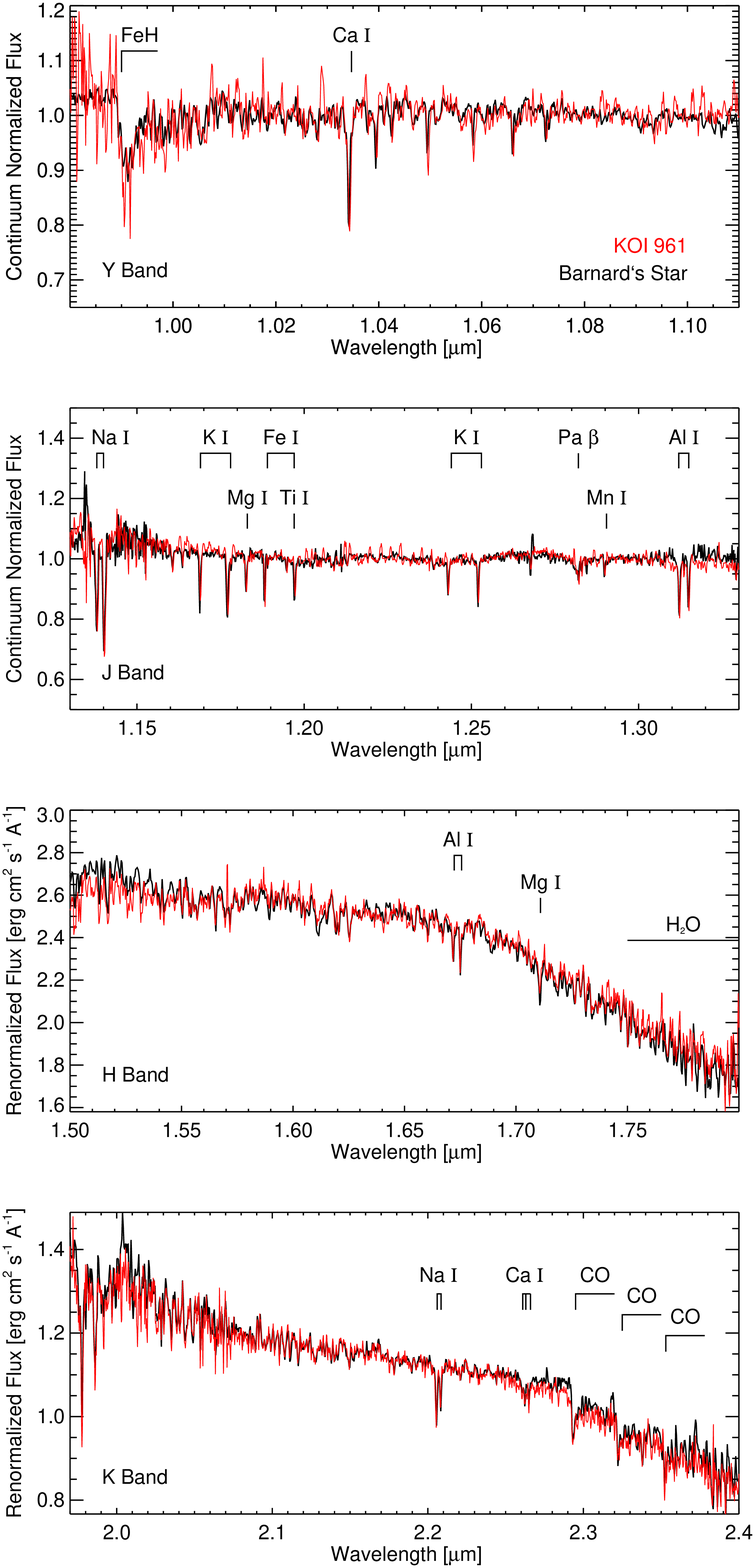}
\caption{Comparison of medium-resolution near-infrared spectra of KOI 961 ({\it red}) and Barnard's Star ({\it black}).  {\it 1st Panel}: $Y$-band TripleSpec spectra of both stars.  {\it 2nd Panel}: $J$-band TripleSpec spectra of both stars.  The $Y$- and $J$-band spectra have each been continuum normalized by a 2nd degree polynomial because of incomplete throughput calibration.  {\it 3rd Panel}: $H$-band spectra of both stars.  {\it 4th Panel}: $K$-band spectra of both stars.  The $H$ and $K$-band spectra of Barnard's Star were each multiplied by a normalization factor to match KOI 961.  Notice the strong agreement in line depths \citep[we identified the lines using][]{Cushing2005}, and the similar spectral shape in $H$ and $K$ bands.  The strong correspondence further reinforces the conclusion that they have similar effective temperatures and metallicities.}
\label{tspec}
\end{center}
\end{figure}

Figure \ref{tspec} shows the near-infrared spectra of KOI 961 and Barnard's Star from the observational programs of Paper 1 and RA11.  Both spectra were obtained with TripleSpec on the Palomar Observatory 200-inch Hale Telescope \citep{Herter2008}.  We reduced the KOI 961 spectrum with the SpexTool program, modified for Palomar TripleSpec \citep[][; M. Cushing, {\it priv. comm.}]{Cushing2004} and used spectra of nearby A0V stars for telluric removal and flux calibration with the {\tt xtellcor} package \citep{Vacca2003}.  The RA11 spectrum was reduced using a custom package written by P.S.M., based on the reduction package for the TEDI instrument, described in \citet{Muirhead2011}.  We supplemented the Barnard's Star spectrum with an additional observation taken on UT 19 October 2011.  The supplemental observation was taken using an observing sequence allowing for more reliable $Y$-band data reduction.

Figure \ref{tspec} shows spectra for $Y$-, $J$-, $H$- and $K$-bands.  The throughput calibration in $Y$ and $J$ bands is incomplete, so we continuum normalized by fitting a quadratic function to the spectrum within each of the bands and then dividing by the fit.  Regions between the bands are heavily obscured by telluric lines.  The strong correspondence between the stars' spectral shapes and line-depths further reinforces the conclusion that they have similar effective temperatures and metallicities.

\subsection{High-Resolution Optical Spectroscopy}

We obtained high-resolution optical spectra of KOI 961 with the HIgh Resolution Echelle Spectrometer (HIRES) on the Keck I Telescope \citep{Vogt1994} on UT 9 October 2011.  We exposed for 1060 seconds using settings that achieved a resolving power of 55000 from 3630 to 8900 $\rm \AA$, and a signal-to-noise ratio of 25 at 6400 $\rm \AA$.

Barnard's Star is currently monitored for orbiting exoplanets as part of the California Planet Search, a precision radial-velocity survey using an iodine absorption cell for calibration \citep[e.g.][]{Johnson2011a}.  We combined all of the spectra taken of Barnard's Star over the course of the survey using regions that are not contaminated by molecular iodine absorption.  The result is a spectrum with an average signal-to-noise ratio of roughly 2500.

\begin{figure*}
\begin{center}
\includegraphics[width=6.5in]{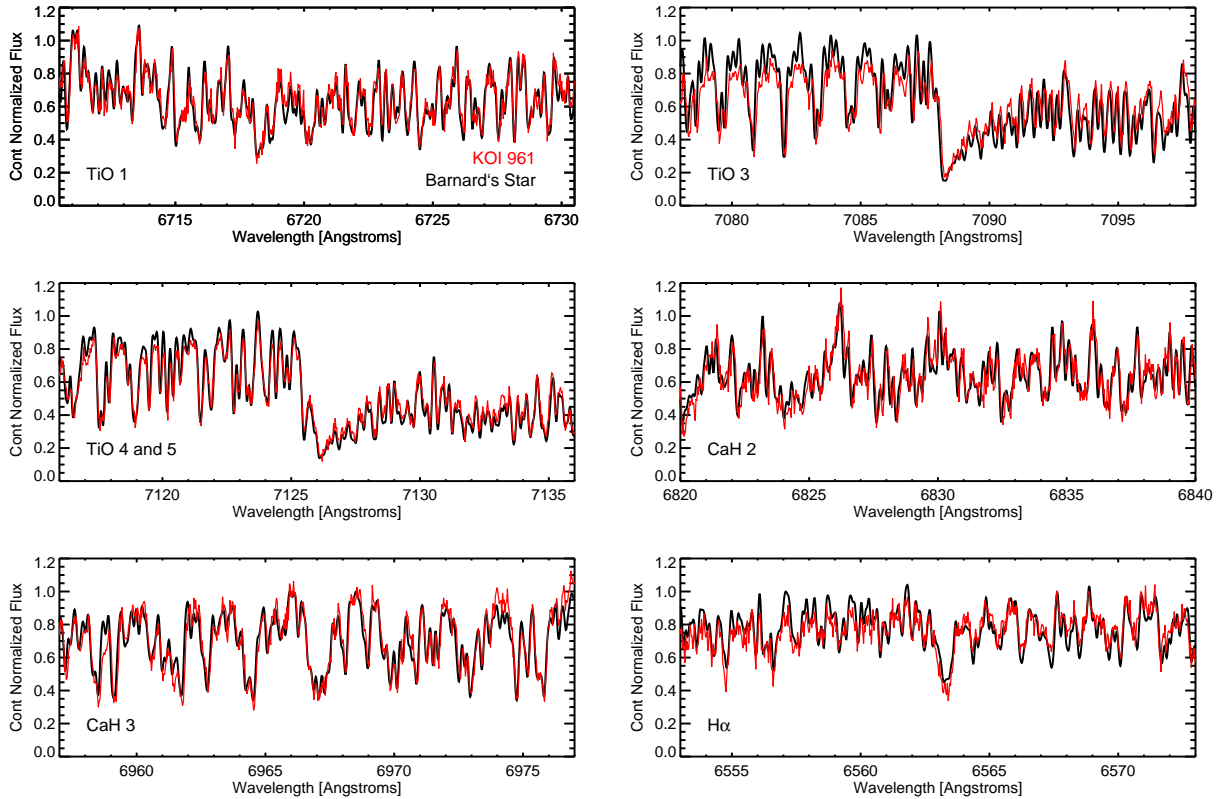}
\caption{Comparison of HIRES spectra of KOI 961 ({\it red}) and Barnard's Star ({\it black}) for select molecular bands identified in \citet{Reid1995} and  H-$\alpha$.  The spectra are continuum normalized within each spectral region.  The spectra are remarkably similar, and neither shows appreciable rotational broadening. The discrepancy in TiO3 ({\it top right}) is attributed to errors in the normalization of the spectra, as the TiO3 region is close to the edge of a HIRES order.  Neither spectrum shows H$\alpha$ emission, which has been shown to correlate with galactic disk height and age for M dwarfs \citep{West2004, West2008}; however, Barnard's Star is known for occasional H$\alpha$ flaring \citep[e.g.][]{Paulson2006}.}
\label{hires}
\end{center}
\end{figure*}

Figure \ref{hires} shows the HIRES spectra of KOI 961 and Barnard's Star in molecular bands corresponding to select PMSU indices, including H$\alpha$.  The correspondence is remarkable; however, unlike the moderate-resolution spectrum in Figure \ref{dbspec}, the HIRES spectra are continuum normalized within each order using a quadratic fit.  This will hide differences in the bulk depths of molecular bands, which are more apparent at moderate resolution.  Nevertheless, the striking similarity attests to the presence of similar molecular opacities, which are a strong function of abundance and temperature in M dwarfs.  Like the moderate-resolution spectra, neither star shows H$\alpha$ emission.  \citet{West2004} and \citet{West2008} showed that the fraction of mid-to-late M dwarfs with H$\alpha$-emission decreases with height above the galactic plane, and the lack of H$\alpha$ in either star indicates that they are part of an old, high disk-scale-height population, consistent with their low-metallicities.  \citet{West2008} estimate the activity lifetime of M4 dwarfs to be 4.5  $\pm^{0.5}_{1.0}$ Gyr, providing a lower limit on the ages of KOI 961 and Barnard's Star.

\subsubsection{Projected Rotational Velocity}

upper limit to the projected rotational velocity of Barnard's star, finding Vsin(i) < 2.5 km/s.

\citet{Browning2010} report upper limit to the projected rotational velocity of Barnard's star, finding $V\sin (i)$ $<$ 2.5 km $\rm s^{-1}$.  This value represents the lowest $V\sin (i)$ measurable given the resolving power of HIRES.  KOI 961 shows no appreciable rotational broadening compared to Barnard's Star, indicating similar slow rotation.  To quantify the rotational broadening, we followed the approach of \citet{Browning2010}.  

\begin{figure}
\begin{center}
\includegraphics[width=3.3in]{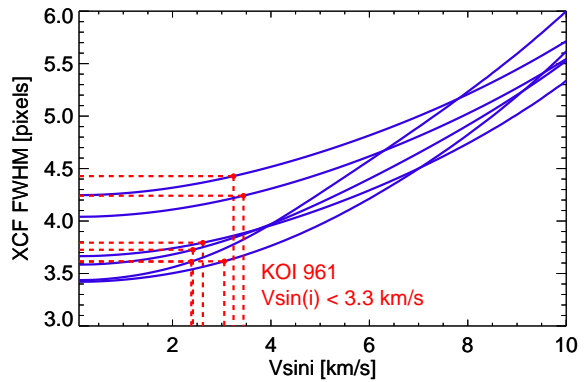}
\caption{Cross-correlation full-width-at-half-maximum (XCF FWHM) vs. $V\sin (i)$.  {\it Solid blue lines}: XCF FWHM for  Barnard's Star cross-correlated with a rotationally-broadened version of itself for each of the six spectral regions in Figure \ref{hires}.  {\it Dashed red lines}:  XCF FWHM for KOI 961 cross-correlated with Barnard's Star, interpolated onto the the blue lines to measure $V\sin (i)$.}
\label{xcf}
\end{center}
\end{figure}

For each of the spectral regions in Figure \ref{hires}, we cross-correlated the spectrum of Barnard's Star with a rotationally-broadened version of itself, using a convolution kernel of specified $V\sin (i)$.  We did this for several $V\sin (i)$ values from 0.1 to 10 km $\rm s^{-1}$.  We fitted Gaussian functions to the central peaks of the resulting cross-correlation functions (XCFs), and recorded the full-width-at-half-maximum (FWHM) of each fit.  Figure \ref{xcf} shows the resulting XCF FWHM vs. $V\sin (i)$ of the rotational kernels for the six spectral regions ({\it solid blue lines}).  This provides a calibration to convert XCF FWHM into $V\sin (i)$ for each spectral region.

We then cross-correlated the spectrum of Barnard's Star with KOI 961, and interpolate that value onto the FWHM-$V\sin (i)$ curves for each spectral region ({\it dashed red lines} in Figure \ref{xcf}).  The mean $V\sin (i)$ for the six regions if 2.9 km $\rm s^{-1}$, and the standard deviation is 0.4 km $\rm s^{-1}$, showing that both Barnard's Star and KOI 961 are slowly rotating.  We caution that this $V\sin (i)$ value should be treated as an upper limit rather than an absolute measurement, since this is near the resolving power of HIRES, and we estimate that upper limit to be one sigma higher than the mean: 3.3 km $\rm s^{-1}$.  

Assuming the inclination of the rotational axis of the star is 90 degress, and the 0.17 $R_\Sun$ stellar radius calculated in the following section, we calculated the rotation period of KOI 961 to be greater than 2.6 days.  Such slow rotation is consistent with KOI 961 being an older mid-M dwarf \citep[$>$ 1-2 Gyr, Figure 12 from][]{Irwin2011a}; however, the low $V\sin (i)$ measurement does not constrain the age as well as the lack of H-$\alpha$ ($>$ 4.5 Gyr), or the kinematic properties (see Section \ref{parameters}).

\subsubsection{Absolute Radial Velocity}

Without the iodine cell, the wavelength-solution of HIRES can be accurately calibrated to 0.1 km $\rm s^{-1}$ using the telluric A and B bands \citep{Chubak2012}.  Using the telluric calibration, we measured the radial velocity difference between KOI 961 and Barnard's Star by cross-correlating the wavelength-calibrated spectrum of KOI 961 with an iodine-free ``template'' HIRES spectrum of Barnard's Star taken as part of the California Planet Search.  Applying the difference to a previously measured absolute radial velocity of Barnard's Star \citep[$\rm V_r$ = -110.85 $\pm$ 0.23 km $\rm s^{-1}$,][]{Marcy1989}, we measured the absolute radial velocity of KOI 961 to be -84.48 $\pm$ 0.2 km $\rm s^{-1}$.  

\section{Estimation of Stellar Parameters for KOI 961}\label{parameters}

The qualitative similarities between KOI 961 and Barnard's Star are striking.  However, to accurately estimate the stellar parameters of KOI 961--in particular mass, radius and luminosity--we must use quantitative, calibrated measurements.  For this purpose, we used the $K$-band metallicity and effective temperature measurements of the two stars from RA11 and Paper 1, which are consistent with the qualitative observations of Section \ref{observables}.   Table \ref{kband} summarizes the $K$-band measurements.

\begin{table}
\begin{center}
\caption{Comparison of Measured Parameters with $K$-band Spectra\label{kband}} 
\begin{tabular}{l c c } 
\hline\hline                        
Parameter & Barnard's Star & KOI 961\\
\hline                  
$T_{\rm eff}$ & 3266 $\pm$ 29 K\footnotemark[1] & 3200 $\pm$ 65 K\\

[M/H] & -0.27 $\pm$ 0.12 & -0.33 $\pm$ 0.12\\

[Fe/H] & -0.39 $\pm$ 0.17 & -0.48 $\pm$ 0.17\\

\hline 
\end{tabular}
\footnotetext[1]{The $K$-band $T_{\rm eff}$ measurements are calibrated using the BT-Settl-2010 model stellar atmospheres \citep{Allard2010}.  The uncertainty in $T_{\rm eff}$ is calculated using Monte Carlo simulations of noise in a measured spectrum.  However, the uncertainty in the calibration itself is not currently included.}
\end{center}
\end{table}

In Paper 1, we interpolated the $K$-band measurements of effective temperature and metallicity onto evolutionary isochrones to estimate the stellar parameters of low-mass KOIs, including KOI 961.  However, empirical measurements of low-mass stellar parameters and predictions from evolutionary isochrones show significant discrepancies, particularly with regard to stellar radius and effective temperature \citep{Ribas2006, Torres2011}.  In this paper, we use evolutionary isochrones to estimate the {\it differences} in stellar parameters between Barnard's Star and KOI 961.  By differencing the interpolated parameters, systematic offsets in the models or measurements are significantly reduced.  We then apply those differences to the {\it empirical} parameters of Barnard's Star to estimate the parameters of KOI 961, and corresponding uncertainties.  By ``empirical,'' we mean measurements that do not depend on stellar evolutionary models.

\subsection{Summary of Empirical Measurements of Barnard's Star}

Unlike most known M dwarfs, Barnard's Star is bright enough to have a reliable parallax measurement from the {\it Hipparcos} satellite \citep{vanleeuwen2007}, providing absolute photometric magnitudes for the bands described in Section \ref{photometry}.  \citet{Delfosse2000} derived mass-luminosity relations for M dwarfs using nearby M-dwarf eclipsing binary systems; of their relations, the $M_{K_{\rm CIT}}$-mass relation has the lowest residuals with the least dependence on stellar metallicity (see Figure 3 from their paper).  We calculated $M_{K_{\rm CIT}}$ by combining the $K_{\rm CIT}$ measurement reported in \citet{Leggett1992} with the {\it Hipparcos} parallax.  Using the empirical $M_{K_{\rm CIT}}$-mass relation of \citet{Delfosse2000}, we calculated a stellar mass of 0.158 $\rm M_\Sun$ for Barnard's Star.

\citet{Delfosse2000} do not report the systematic uncertainty in the $M_{K_{\rm CIT}}$-mass relation.  To estimate the uncertainty in the relation, we calculated the root-mean-square ($RMS$) of the residuals between $M_{K_{\rm CIT}}$ of the calibration stars and those of the relation, and found it to be 0.18 magnitudes.  This is much larger than the uncertainty in $M_{K_{\rm CIT}}$ of Barnard's Star, and will dominate the uncertainty in its measured mass.

We estimate the uncertainty in the mass of Barnard's Star using a Monte Carlo approach.  We recalculated the mass of Barnard's Star for 500 instances of $M_{K_{\rm CIT}}$, each with noise added, drawn from a Gaussian distribution of standard deviation of 0.18 magnitudes.  The resulting standard deviation of the mass distribution is 0.013 $\rm M_\Sun$.

The radius of Barnard's Star has been measured previously using optical long-baseline interferometry.  \citet{Lane2001} reported a radius of 0.201 $\pm$ 0.008 $R_\Sun$ for Barnard's Star, using observations with the Palomar Testbed Interferometer.  The uncertainty in the stellar radius comes from a combination of measurement errors, uncertainty in the angular size of the calibrator stars and uncertainties in the adopted limb-darkening parameters.  \citet{Segransan2003} recalculated the radius using a different set of limb-darkening coefficients, and reported a radius of 0.196 $\pm$ 0.008 $\rm R_\sun$ for Barnard's Star, consistent with \citet{Lane2001} to within the quoted uncertainty.  This shows that the interferometrically measured stellar diameter is robust against assumptions about the limb darkening and strengthens our confidence that this radius measurement represents an accurate measurement of the true radius of Barnard's star.

Finally, \citet{Dawson2004} combined and calibrated available spectra of Barnard's Star across optical and infrared wavelengths to accurately measure its luminosity [(3.46 $\pm$ 0.17) $\times$ $10^{-3}$ $\rm L_\sun$], which, when combined with the radius measurement, provides an empirical measurement of the star's effective temperature (3134 $\pm$ 102 K).  This agrees remarkably well with the effective temperature measured using $K$-band in RA11 of 3266 $\pm$ 29 K.  The quoted uncertainty for the effective temperature measured in RA11 only accounts for measurement errors in the $K$-band spectrum, and not systematic effects from errors in the calibration.

\subsection{Perturbation Analysis}

To estimate the {\it difference} in mass, radius and luminosity between Barnard's Star and KOI 961, we interpolated the $K$-band measurements of stellar effective temperature and metallicity onto the 5-Gyr isochrones of \citet{Baraffe1998}.  We use the \citet{Baraffe1998} models because, unlike other sets of evolutionary isochrones, they include grid points with stellar effective temperatures below 3200 K for sub-solar metallicity.  The \citet{Baraffe1998} isochrones are available in two metallicities, [M/H] = 0.0 and -0.5, which bracket the measured metallicities of Barnard's Star and KOI 961.  The [M/H] = -0.5 isochrones are only available for a mixing-length parameter ($\alpha$) of 1 and for a Helium abundance ([Y]) of 0.25, so we choose the [M/H] = 0.0 isochrones with those same parameters.  To test the effect of age on the parameter estimation, we repeated the calculations using 10-Gyr isochrones.  The resulting stellar parameters were only marginally different compared to the uncertainties, a consequence of the slow evolution of M dwarfs.

\begin{figure}
\begin{center}

\includegraphics[width=3.3in]{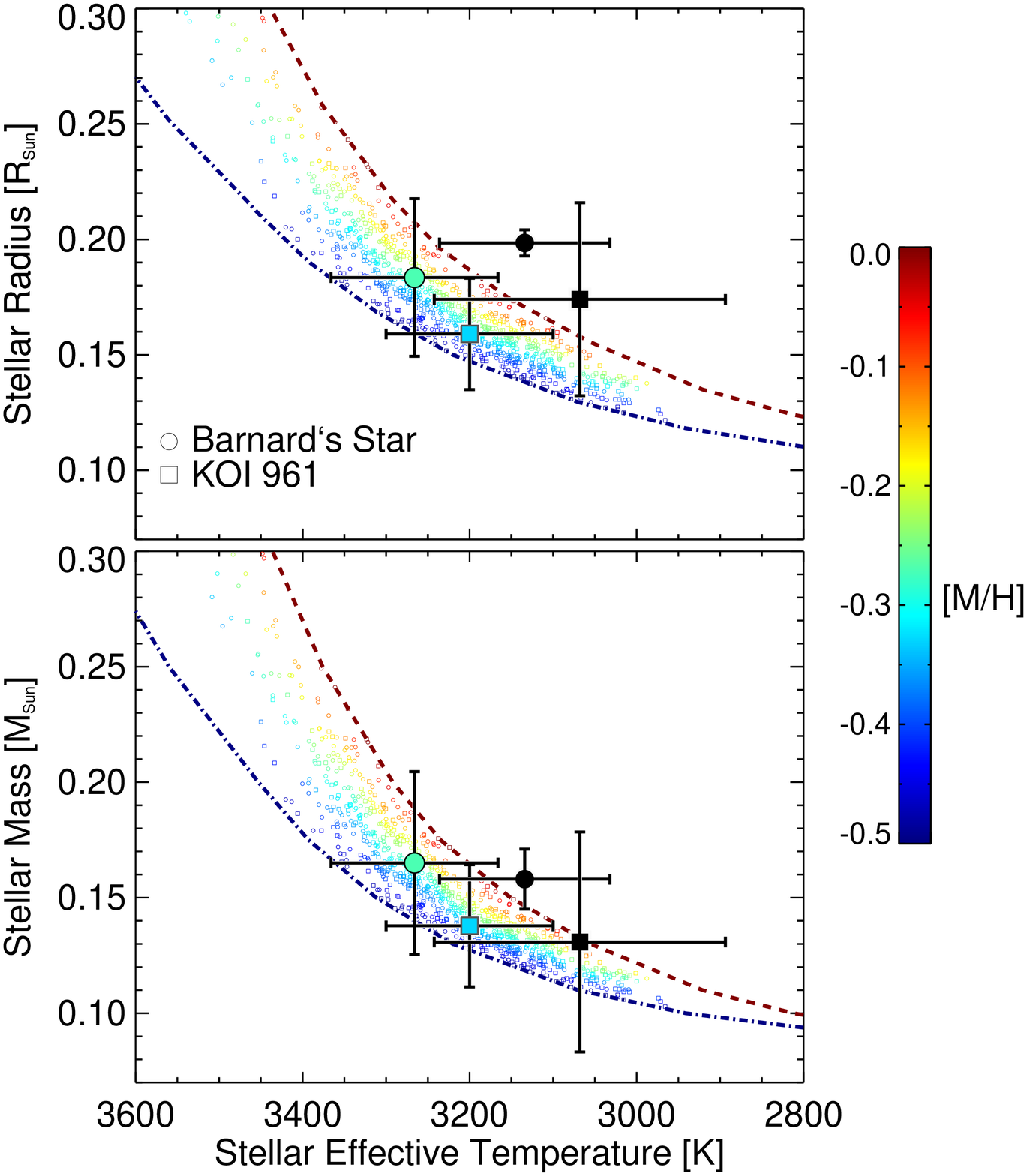}
\caption{{\it Top Panel}: Stellar radius vs. effective temperature.  {\it Bottom Panel}: Stellar mass vs. effective temperature.  Colors denote metallity, [M/H].  {\it Lines}: The 5-Gyr evolutionary isochrones of \citet[][]{Baraffe1998} for [M/H] = 0.0 and [M/H] = -0.5.  {\it Large colored circle and square}: Interpolated position of Barnard's Stars and KOI 961, respectively, based on $K$-band measurements of metallicity and effective temperature.  {\it Small colored circles and squares}: Monte Carlo simulations of noise in the $K$-band measurements, used to estimate the error in the mass, radius and luminosity of the interpolated values.  {\it Black circle}: Empirically measured values for Barnard's Star and uncertainties.  {\it Black square}: Our estimate for the radius and effective temperature of KOI 961, found by applying the difference in interpolated values to the empirical measurements of Barnard's Star.}
\label{interpolate}
\end{center}
\end{figure}

Figure \ref{interpolate} shows stellar radius and stellar mass vs. effective temperature for the 5-Gyr isochrones of \citet{Baraffe1998} and the interpolated location of Barnard's Star and KOI 961 based on the $K$-band measurements.  We applied the difference in mass, radius and luminosity between the model interpolants to the empirical measurements of Barnard's Star to estimate the values for KOI 961.

To estimate the uncertainties in KOI 961's parameters, we interpolated onto the isochrones 500 Monte Carlo simulations of the $K$-band $T_{\rm eff}$ and [M/H] measurements, each with random noise added.  The noise is drawn from a normal distribution with standard deviation of 100 K for $T_{\rm eff}$ and 0.12 dex for [M/H].  We then calculated the standard deviation of the distribution in the resultant radii, masses and luminosities.  We chose to inflate the measurement uncertainties in $T_{\rm eff}$ to 100 K to account for potential errors in the calibration.  For [M/H] we used an uncertainty of 0.12 based on the calibration uncertainty calculated in \citet{Rojas2010}.  This provides uncertainties in the {\it interpolated} masses, radii and luminosities of KOI 961 and Barnard's Star, which, when added in quadrature, provide the uncertainty in the difference between the stars' parameters.  We then added in quadrature the uncertainties in the differences to the empirical measurements of Barnard's Star, which provides the uncertainty of KOI 961's parameters.  Table \ref{barnards_table} lists the empirical measurements of Barnard's Star and our estimated values for the stellar parameters of KOI 961 and their uncertainties.  

\begin{table}
\begin{center}
\caption{Barnard's Star and KOI 961 Stellar Parameters} 
\begin{tabular}{l c c } 
\hline\hline                        
Parameter & Barnard's Star & KOI 961\\ [0.5ex] 
 & (empirically measured) & (this work)\\ [0.5ex] 
\hline                  
$R_\star$ & 0.199 $\pm$ 0.006  $\rm R_\sun$ \footnotemark[1] & 0.17 $\pm$ 0.04 $\rm R_\sun$ \\
$M_\star$ & 0.158 $\pm$ 0.013 $\rm M_\sun$\footnotemark[2] & 0.13 $\pm$ 0.05 $\rm M_\sun$ \\ 
$L_\star$ & (3.46 $\pm$ 0.17) $\times$ $10^{-3}$ $\rm L_\sun$\footnotemark[3] & 2.40  $\times$ $10^{-3.0 \pm 0.3 }$ $\rm L_\sun$\footnotemark[4] \\
$T_{\rm eff}$ & 3134 $\pm$ 102 K\footnotemark[3] & 3068 $\pm$ 174 K \\ 
$M_{K, \rm CIT}$ & 8.21 $\pm$ 0.03\footnotemark[5,6] &  8.55 $\pm$ 0.35\\ 
$M_{K_{\rm S}}$ & 8.22 $\pm$ 0.03\footnotemark[5,7]  &  8.53 $\pm$ 0.35 \\ 
$d$ & 1.824 $\pm$ 0.005 pc \footnotemark[7] & 38.7 $\pm$ 6.3 pc\\
U & -134 $\pm$ 1 km $\rm s^{-1}$ & 50 $\pm$10 km $\rm s^{-1}$\\
V & 18 $\pm$ 1 km $\rm s^{-1}$ & -70 $\pm$ 1 km $\rm s^{-1}$\\
W & 27 $\pm$ 1 km $\rm s^{-1}$ &  -59 $\pm$ 8 km $\rm s^{-1}$\\
$V\sin(i)$ & $<$ 2.5 km $\rm s^{-1}$\footnotemark[8] & $<$ 3.3 km $\rm s^{-1}$\\
\hline 
\end{tabular}
\footnotetext[1]{Weighed mean of \citet[][0.201 $\pm$ 0.008 $\rm R_\sun$]{Lane2001} and \citet[][0.196 $\pm$ 0.008 $\rm R_\sun$]{Segransan2003}}
\footnotetext[2]{Calculated using the $M_{K, \rm CIT}$-mass relations of \citet{Delfosse2000}.}
\footnotetext[3]{\citet{Dawson2004}}
\footnotetext[4]{Since luminosity is a strong power of the effective temperature and radius, the distribution of interpolated Monte Carlo luminosities is  asymmetric.  Therefore, we quote the error in the inferred luminosity of KOI 961 in the exponent, equivalent to the standard deviation of the log of the resultant luminosity distribution.}
\footnotetext[5]{Parallax from \citet{vanleeuwen2007}}
\footnotetext[6]{\citet{Leggett1992}}
\footnotetext[7]{2MASS, \citet{Cutri2003}}
\footnotetext[8]{\citet{Browning2010}}
\label{barnards_table} 
\end{center}
\end{table}

\begin{figure}
\begin{center}
\includegraphics[width=3.3in]{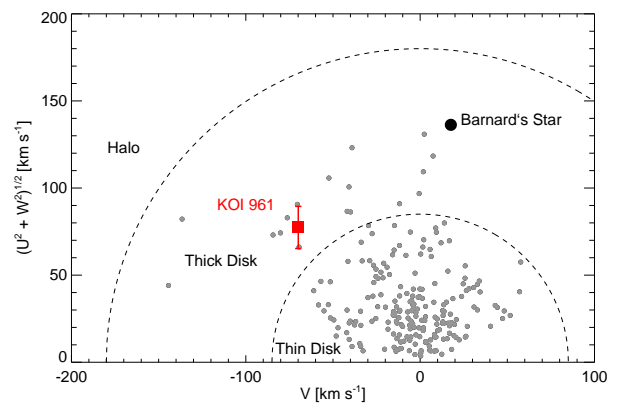}
\caption{Toomre Diagram of G- K- and M-type primary stars with measured trigonometric parallaxes $>$ 100 $mas$.  We include the thin and thick disk boundaries defined by \citet{Fuhrmann2004}.  Barnard's Star ({\it red circle}) and KOI 961 ({\it blue square}) are indicated and have space velocities consistent with thick-disk stars.  We corrected for the solar motion using the values of \citet{Francis2009}: U = 7.5 + -1.0, V = 13.5 + 0.3, W = 6.8 + -0.1 km/s.}
\label{toomre}
\end{center}
\end{figure}

We calculated the space motion of KOI 961 by combining the proper motion measurements of \citet{Lepine2005} with our estimated distance and measured radial velocity.  Figure \ref{toomre} shows a Toomre diagram of nearby G- K- and M-type primary stars with measured trigonometric parallaxes greater than 100 $mas$.  Using the population definitions of \citet{Fuhrmann2004}, it is clear that both Barnard's Star and KOI 961 have kinematic properties consistent with the ``thick disk'', containing stars with ages of 7 to 13 Gyr \citep{Feltzing2008}.  We acknowledge that the presence of a thick disk in the Milky Way is controversial, however the difference in stellar kinematics as a function of disk scale-height the galactic disk is firmly established \citep[e.g.][]{Bovy2011}.  The kinematic properties, metallicities, H$\alpha$ emission and $Vsin(i)$ measurements of KOI 961 and Barnard's Star are all consistent with an old, high disk-scale-height stellar population.

\subsection{Consistency Check on GJ 1214}

As a consistency check of our methods, we performed an identical perturbation analysis on GJ 1214, a low-mass star with a transiting Super-Earth: GJ 1214 b \citep{Charbonneau2009}.  \citet{Carter2011} measured multiple transit light-curves of GJ 1214 b and calculated the mean density of the star using the fitted transit parameters \citep[e.g.][]{Seager2003}.  GJ 1214 has a parallax measurement by \citet{vanaltena1995} and 2MASS $JHK$ photometry.  Combining the mass-$M_K$ relation of \citet{Delfosse2000} with the transit-derived stellar density, \citet{Carter2011} reported a stellar mass of 0.157 $\pm$ 0.012 $\rm M_\Sun$ and a stellar radius of 0.210 $\pm$ 0.007 $\rm R_\Sun$ for GJ 1214 ({\it Method A} from their paper).  The stellar radius was larger than estimated using photometric relationships from the \citet{Baraffe1998} evolutionary isochrones, which they also report as $R_\star$ = 0.179 $\pm$ 0.006 $\rm R_\Sun$ ({\it Method B} from their paper).

RA11 report a stellar effective temperature of 3245 $\pm$ 31 K and metallicity ([M/H]) of 0.15 $\pm$ 0.12 for GJ 1214 using the $K$-band spectroscopic indices.  With an effective temperature similar to Barnard's Star and KOI 961, as well as available $K$-band measurements, GJ 1214 provides a valuable opportunity to test our perturbation analysis.\footnote{Other possible test stars include the short-period M dwarf eclipsing-binaries CM Draconis and KOI 126, which have similar effective temperatures to Barnard's Star and KOI 961 ($\sim$ 3000 K) and highly accurate mass and radius measurements \citep[][]{Lacy1977, Metcalfe1996, Morales2009, Carter2011b}.  However, it has been suggested that mutual magnetic interactions in short-period M dwarf binaries  introduce discrepancies in the mass-radius-effective temperature relationships as compared to single field M dwarfs \citep{Chabrier2007}, and recent observational evidence supports this scenario \citep{Kraus2011}.  Also, measuring the $K$-band $T_{\rm eff}$ and [M/H] indices of individual stars in unresolved, short-period eclipsing binaries is challenging and has not yet been done.  Therefore, we believe GJ 1214 is the best test object.}

We proceed assuming Solar-metallicity for GJ 1214, since we cannot interpolate the super-Solar $K$-band value of [M/H] = +0.15 onto the isochrones of \citet{Baraffe1998}.  Using the same perturbation procedure used for KOI 961, and the $K$-band effective temperature measurement, we calculate a a stellar mass of 0.17 $\pm$ 0.05 $\rm M_\Sun$ and a stellar radius of 0.21 $\pm$ 0.04 $\rm R_\Sun$ for GJ 1214.  The values are remarkably consistent with the results of {\it Method A} by \citet{Carter2011} to within our estimate uncertainties.  Incorporating the $K$-band metallicity measurement of GJ 1214 would result in a larger radius had we used super-Solar metallicity isochrones, but not by greater than $\Delta R_\star$ = 0.1 $\rm R_\Sun$ (see Figure \ref{interpolate}): still well within the estimated uncertainty.  We conclude that the perturbation analysis is an effective method for determining accurate stellar parameters of M dwarfs in the absence of parallax measurements.

\section{Transit Parameters}\label{transit}

Now we turn to a discussion of the transits detected around KOI 961.  \citet{Borucki2011} reported transit parameters (period, duration, impact parameter and planet-star radius ratio) for the three planet-candidates orbiting KOI 961.  However, the published radius and impact parameter for KOI 961.02 are inconsistent: the impact parameter and planet-to-star radius ratio indicate that the transit should not occur ($b/R_\star$=1.29, $R_P/R_\star$=0.194).  With this in mind, we chose to perform an independent fit to the public {\it Kepler} data for the three planet-candidates.  In the following section, we use the resulting parameters to calculate the probability that the transit signals are not planetary in nature, and we find this probability is less than 1\% for all three signals.  Therefore, in this section, we refer to them as planets rather than candidates.

We use Levenberg-Marquardt least-square minimization to refit the
light curve of KOI 961 and determine revised transit parameters for
all three planets.  This task is complicated by a number of
simultaneous transits of two or more planets.  For these
situations, we assume that the total loss of light is the
superposition of the losses computed individually for each planet.
We do not attempt to model the short brightening in the light curve as
a result of the possible mutual alignment of three or more bodies [as
described by \citep{Ragozzine2010}].

In principle, one could describe the transit model for each planet
with three minimal parameters: the planet to star radius ratio, the
orbital inclination, and the instantaneous separation of planet and
star, normalized to the stellar radius: $a/R_\star$.  Additionally, we may specify
the ephemeris and period for each planet and parameters describing
the radial brightness profile of the star.  In the case of KOI-961,
each individual transit duration is comparable to the integration time
for the long cadence {\it Kepler} data.  As a result, there exists a near
exact degeneracy between these three minimal parameters.  Fortunately,
the normalized instantaneous separation can be constrained given our
knowledge of the star if we are willing to assume the orbits of all
three planets are circular. In this case, $a/R_\star$ is
specified in terms of the mean stellar density and the orbital
period.  For our fitted transit models, we assume circular orbits and apply the mean stellar density calculated from the stellar mass and radius determinations in Section \ref{parameters}.

\begin{figure}
\begin{center}
\includegraphics[width=2.5in]{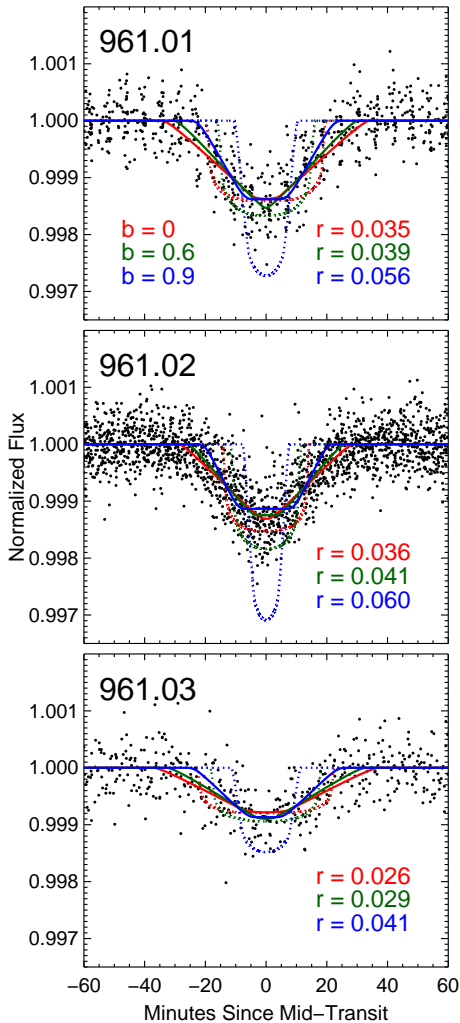}
\caption{{\it Kepler} light curves for each planet after folding the data on the best-fitting linear ephemeris and after having removed the best-fitting transit profiles of the remaining two planets.  We include representative best-fitting light curves -- both the convolved and the continuous profile ({\it solid lines} and {\it dotted lines}, respectively) -- for three impact parameters:  0 ({\it red}), 0.6 ({\it green}) and 0.9 ({\it blue}).  The corresponding, best-fitting radius ratios for each impact parameter are listed in each panel.  High impact parameter solutions ($\gtrsim 1$), as listed for KOI-961 in \citet{Borucki2011}, are disfavored by the data under these constraints, evidenced by the poor match between the solid blue curve and the {\it Kepler} data at the start and end of each folded light curve.}
\label{transits}
\end{center}
\end{figure}

\begin{figure}[]
\begin{center}
\includegraphics[width=2.5in]{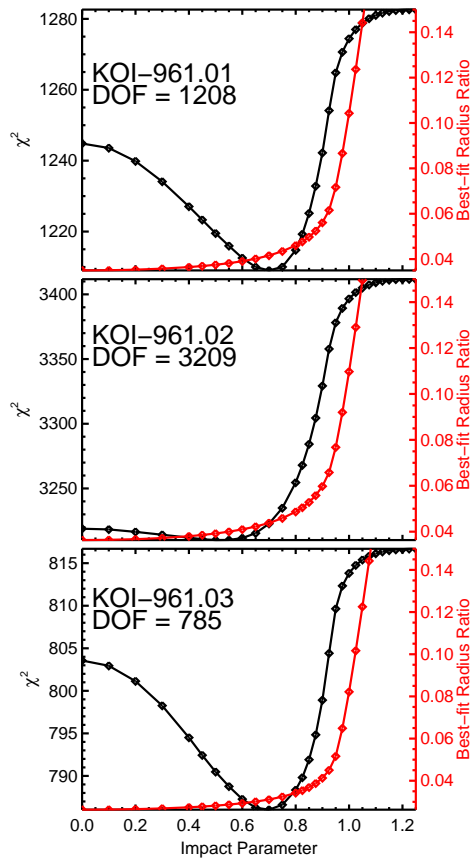}
\caption{The measured fitting statistic, $\chi^2$,
versus the modeled impact parameter for each planet ({\it black)}. In all cases large impact parameters are disfavored by the {\emph Kepler} photometry, and in the case of planets .01 and .03, the light curves provide strong lower constraints on the impact parameter.  We include the best-fitting planet-to-star radius ratio for each impact parameter ({\it red}), illustrating the small spread in radius ratios across the minimum of the $\chi^2$ vs. impact parameter curve.  The best-fitting impact parameters and radius ratios are listed in Table \ref{planet_table}.}
\label{chi2}
\end{center}
\end{figure}

We report in Table \ref{planet_table} the best-fitting radius ratios and
their errors along with the best-fitting impact parameter subject to the
assumption of circular orbits and the specified mean stellar density.  In Figure \ref{transits}, we show the {\it Kepler}
light curve for each planet after folding the data on the best-fitting
linear ephemeris and after having removed the transit profiles of the
remaining two planets.  We also plot representative best-fitting
transit models -- both the convolved and the continuous profile -- for a
few impact parameters.  The corresponding, best-fitting radius ratios
for each impact parameter are listed in each panel.  High impact
parameter solutions ($\gtrsim 1$), as listed for KOI-961 in
\citet{Borucki2011}, are disfavored by the data under these
constraints for all three planets.  As a result, we determine a significantly smaller radius
ratio for all three planets than previously reported.

Combining the best fit transit parameters with our new stellar parameters, we report the planets' physical radii, equilibrium temperatures and corresponding uncertainties in Table \ref{planet_table}.  All three planets have radii smaller than the Earth's, and KOI 961.03 is 1.07 $\pm$ 0.33 times the radius of Mars.  The uncertainties in the planet radii are dominated by the uncertainties in the stellar radius.  To calculate the equilibrium temperatures of the planets, we scaled the equilibrium temperature of Earth at 255 K--which assumes a 30\% albedo and 100\% re-radiation fraction--to the semi-major axes of the planetary orbits and the luminosity of KOI 961 and propagated the uncertainties in those parameters.

\begin{table*}
\begin{center}
\caption{KOI 961 Transit Parameters and Planet Parameters} 
\tabletypesize{\scriptsize}
\begin{tabular}{cccccccc} 
\hline\hline                        

%
KOI & 
$t_0$ [BJD - 2454900] &
Period [days] & 
$a$ [AU] & 
$R_P / R_\star$ & 
 $b$ $[R_\star]$ &
$R_P$ [$\rm R_\Earth$] & 
$T_{\rm eq}$ [K] \\
\hline

961.01& 103.48329 $\pm$ 5.4$\times 10^{-4}$ & 1.2137672 $\pm$ 4.6 $\times 10^{-6}$ & 0.0116 & 0.0419 $\pm$ 0.0018 &    0.711 $\pm$ 0.049 &  0.78 $\pm$ 0.22 & 519 $\pm$ 52\\
961.02 & 66.86952 $\pm$ 4.2$\times 10^{-4}$ & 0.45328509 $\pm$ 9.7 $\times 10^{-7}$ & 0.0060 & 0.0395 $\pm$ 0.0011 &    0.520 $\pm$  0.068 &   0.73  $\pm$ 0.20 & 720 $\pm$  73\\
961.03 & 66.7847 $\pm$ 1.3$\times 10^{-3}$ & 1.865169 $\pm$ 1.4$\times 10^{-5}$ & 0.0154 & 0.0307 $\pm$ 0.0019 &  0.681  $\pm$  0.085 & 0.57 $\pm$ 0.18 & 450 $\pm$ 45\\
\hline 
\end{tabular}
\end{center}
\label{planet_table}
\end{table*}

\section{The Planetary Nature of the Transits}\label{planets}

It is conventional to call transit signals ÒcandidatesÓ until the planets are dynamically confirmed, either by radial velocity or transit timing measurements. However, we do not have Doppler spectroscopy for KOI 961 and the transit signals show no obvious timing variations.  Fortunately, given that there are three transit signals, each with a period of less than 2 days, it would require an extremely contrived scenario to account for all of them by non-planetary means.  In the following section we quantify the probability that each of the signals may be an astrophysical false positive, demonstrating that this probability is low enough to consider all three planets validated.  We also discuss the constraints on potential diluting third light in the aperture, which would lead to underestimating the planetary radii.

\subsection{Observational constraints on false positive scenarios}

Photometric signals that appear to be transiting planets may be caused by astrophysical scenarios other than a planet orbiting the target star.  In this work, we consider a false positive to be any scenario different from a planet transiting a star in the KOI 961 system.  This may be either an eclipsing binary or a transiting planet around a chance-aligned background star or an eclipsing binary that is physically bound to KOI 961 in a hierarchical triple system.  Each of these scenarios involves the presence of another star within the {\it Kepler} aperture of KOI 961. 

\begin{figure}
\begin{center}
\subfigure[POSS-I blue, 1951]{
\includegraphics[width=1.5in]{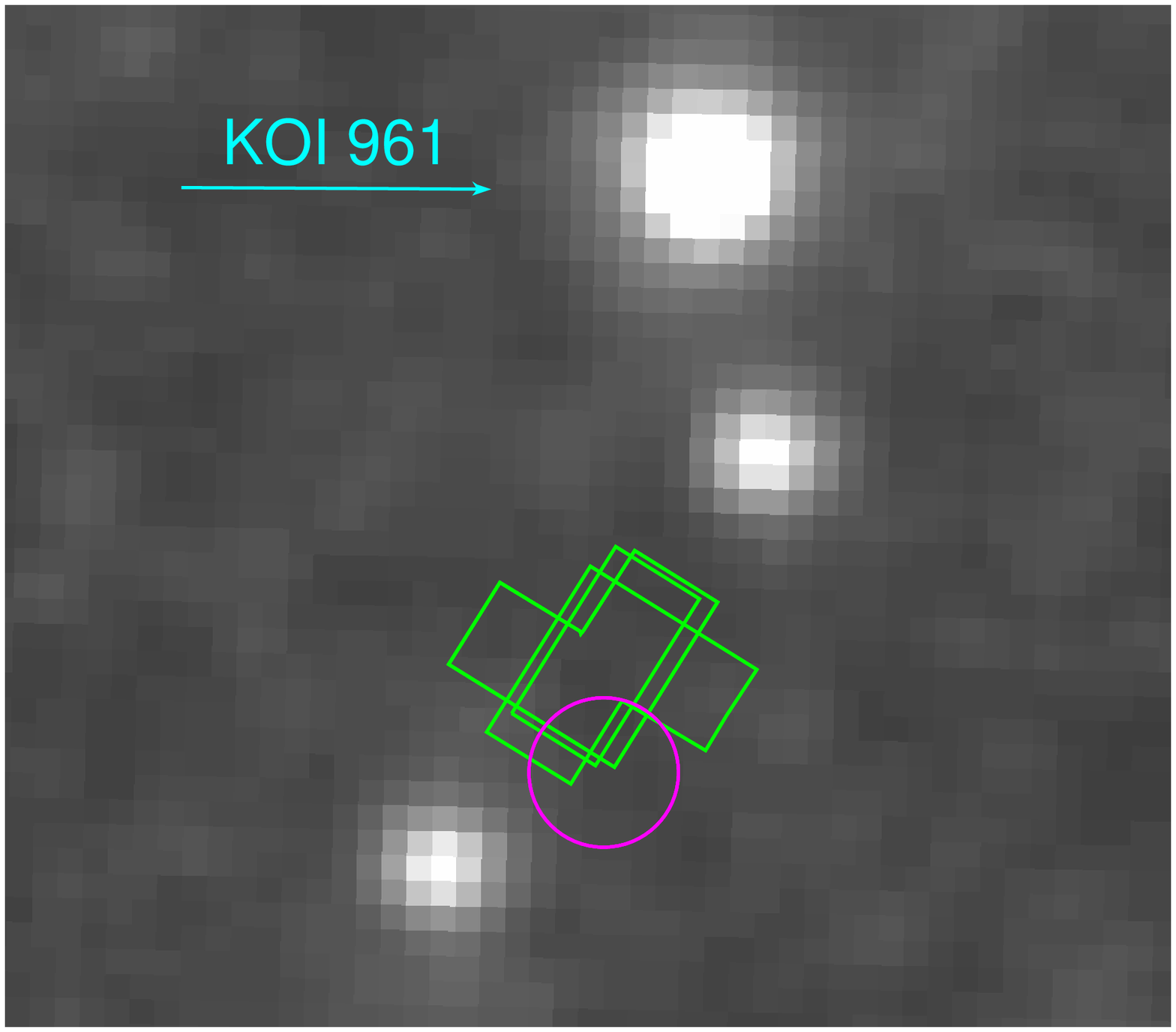}}
\subfigure[POSS-I red, 1951]{
\includegraphics[width=1.5in]{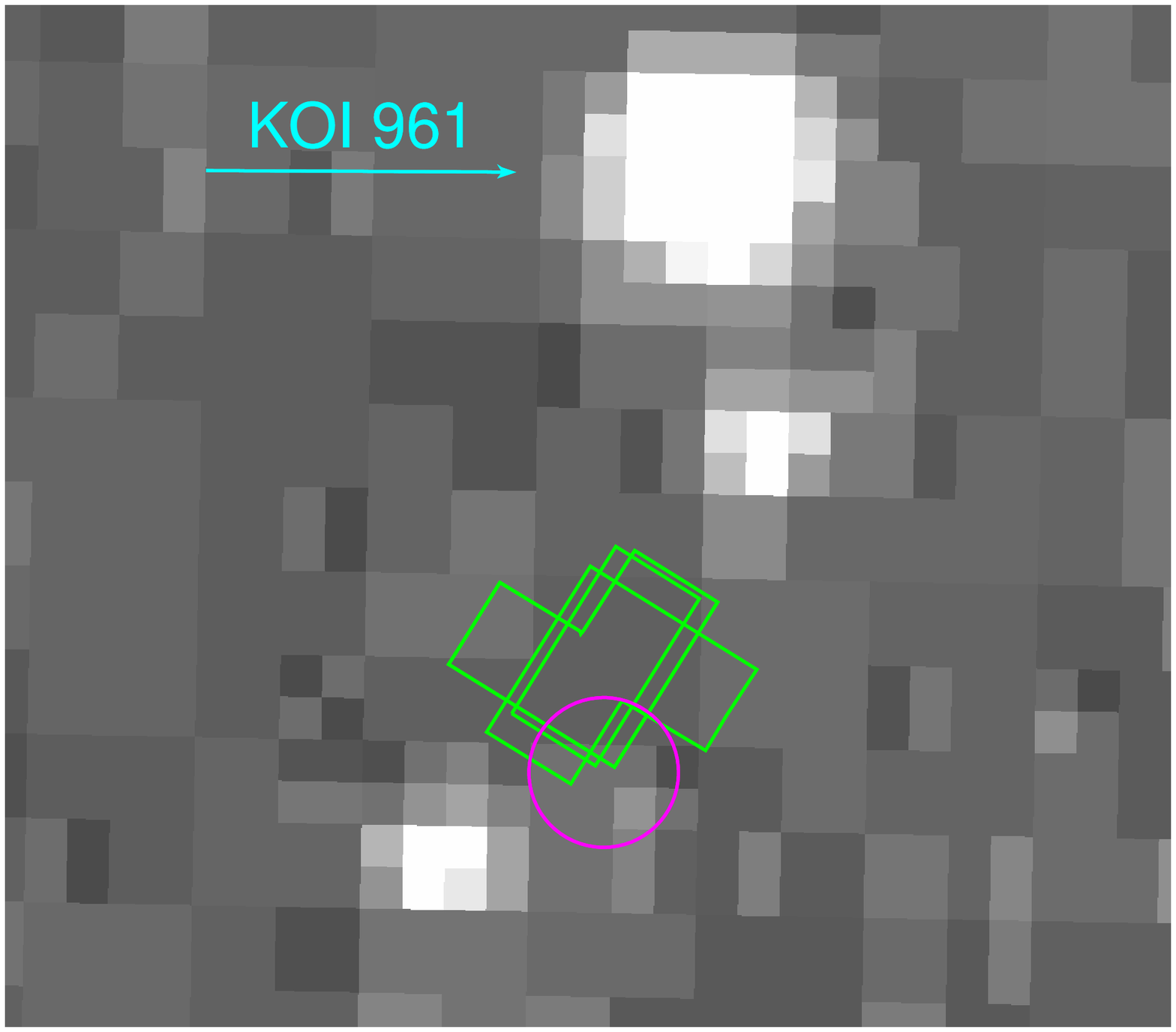}}
\subfigure[POSS-II blue, 1995]{
\includegraphics[width=1.5in]{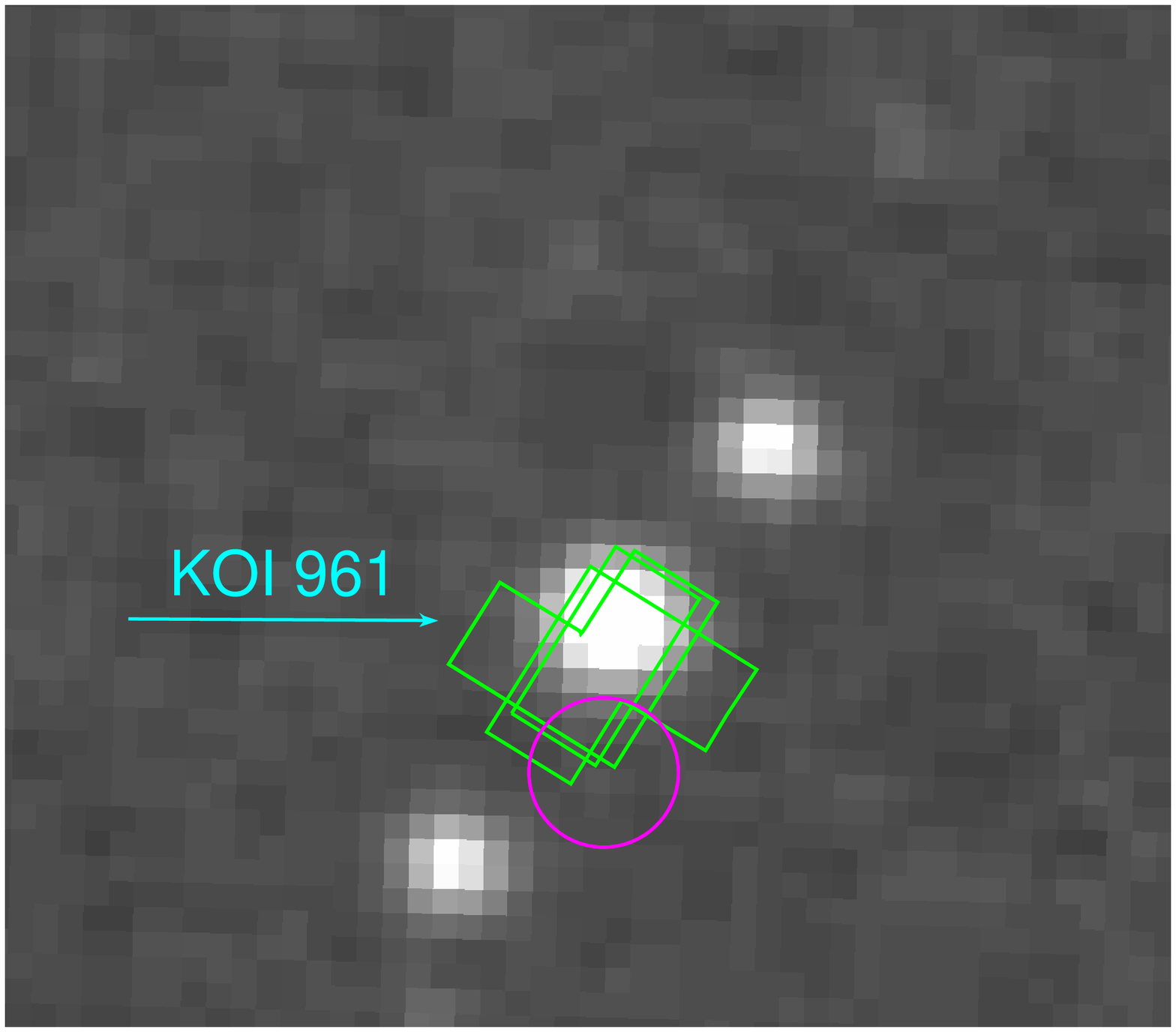}}
\subfigure[POSS-II red, 1991]{
\includegraphics[width=1.5in]{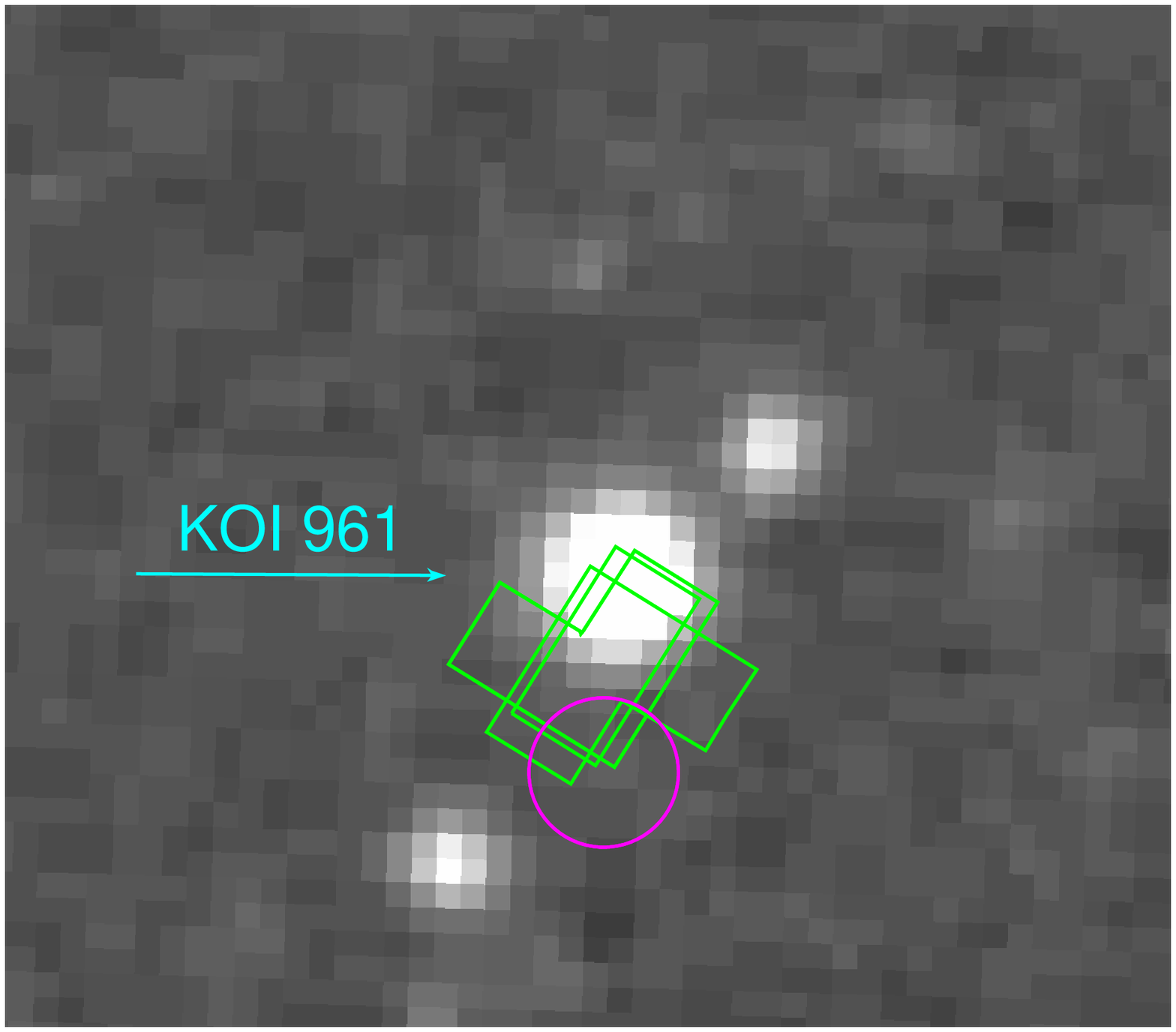}}
\subfigure[HIRES Slit Viewer (this work), 2011]{
\includegraphics[width=1.5in]{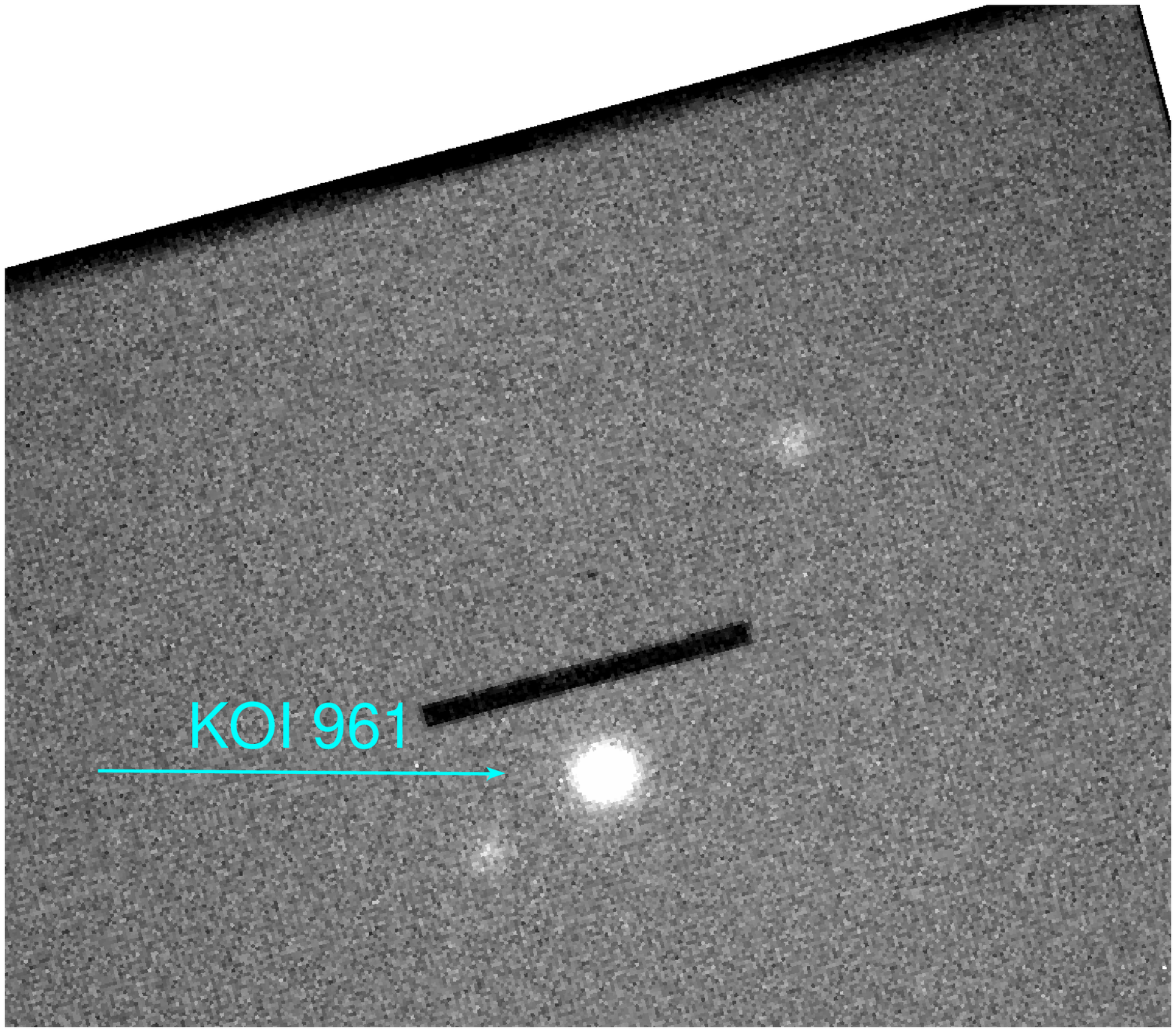}}
\subfigure[HIRES Slit Viewer close-up of KOI 961]{
\includegraphics[width=1.5in]{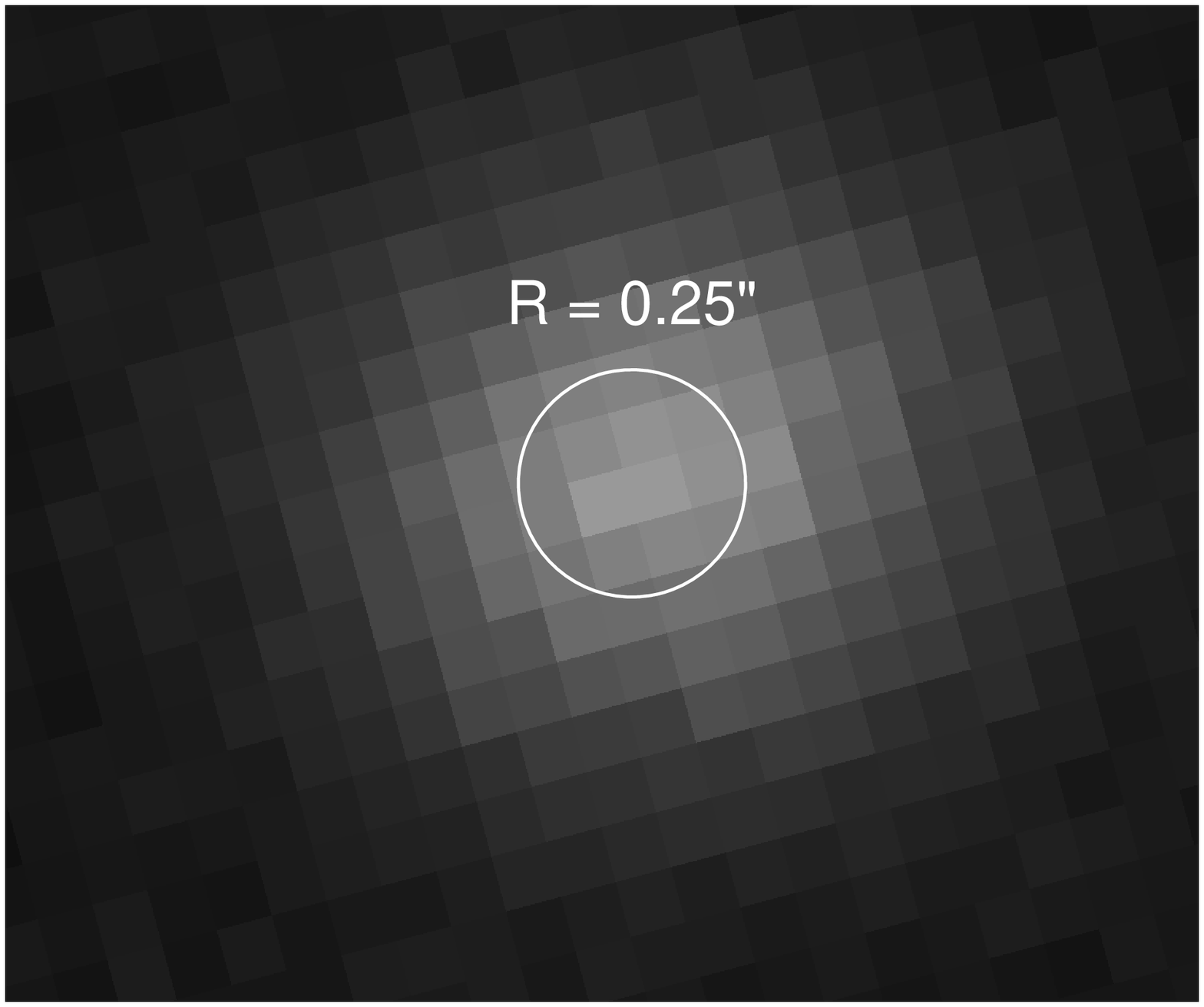}}
\caption{\label{POSS} Images of KOI 961 from the Palomar Observatory Sky Survey \citep[POSS, panels {\it a, b, c} and {\it d}, e.g.][]{Minkowski1963,Reid1991} and from the HIRES slit-viewer (this work, panels {\it e} and {\it f}).  Panels {\it a} through {\it e} are roughly 1 arcminute on each side and show the large proper motion of KOI 961 over 60 years.  The {\it magenta circles} indicate the current (2011) position of KOI 961 and have diameters of 6$\arcsec$, roughly equal to the point-spread-function of {\it Kepler} \citep{Bryson2010, Koch2010}.  We also include the apertures used by {\it Kepler} for flux extraction of KOI 961 during quarters 1, 2 and 3, which we obtained from the Multimission Archive at the Space Telescope Science Institute ({\it green outlines}).  We rule out the presence of a contaminating star $<$ 4.8 magnitudes fainter than KOI 961 using the POSS 1 blue-sensitive image.  We use the POSS images and HIRES slit-viewer image to limit potential false-positive scenarios for the transit signals.\label{poss}}
\end{center}
\end{figure}

We constrain the properties of such a star using archival and recently acquired imaging data, which we show in Figure \ref{poss}.  The large proper motion of KOI 961 allows us to use digitized images from the Palomar Observatory Sky Survey \citep[POSS;][]{Minkowski1963,Reid1991} to rule out the presence of any contaminating object up to 4.8 magnitudes fainter than KOI 961 in the blue-sensitive plates, anywhere within the {\it Kepler} apertures used in the \citet{Borucki2011} data release.  We also use both publicly available $J$-band UKIRT data \citep{Lawrence2007} and a HIRES guider camera image we obtained using Keck I (equivalent of combined $R$- and $V$-band response), which allows us to constrain the presence of blends within a few arcseconds of the current position of KOI 961.

In addition, the HIRES spectrum of KOI 961 constrains the presence of any potentially blending companion through the absence of noticeably contaminating lines.  The symmetry of the cross-correlation profile of the spectrum with that of Barnard's Star reveals that KOI 961 has no companion of comparable brightness ($\Delta V$$\sim$1; or mass ratio $q \sim 0.75$) with velocity separation $>$2 km $\rm s^{-1}$ \citep[e.g.][]{Howard2010}.\footnote{Incidentally, this observational constraint also eliminates the false positive scenario (not discussed in the text) in which one of the transit signals is caused by a similar-mass binary companion to KOI 961 with a grazing eclipse.}  

Using the methods of \citet{Morton2011}, which use simulations of Galactic structure and stellar evolution to estimate the sky density of background stars, we determine that there is a probability of about 0.13 for there to be a chance-aligned star undetected by the POSS images or the Keck guider image anywhere within the {\it Kepler} apertures used for the \citet{Borucki2011} data release.  To determine the extent to which the above observations constrain hierarchical triple scenarios, we simulate a population of stellar binary companions with mass ratio, period, and eccentricity distributions typical of local multiplicity surveys \citep{Raghavan2010,Tokovinin2010}, and assign magnitudes to each companion based on the models of \citet{Baraffe1998}.  We then place each simulated companion at a random point on its orbit (using a uniform distribution of mean anomaly) and use the resulting projected separation (assuming a distance of 40 pc), relative velocity, and magnitude difference to determine whether it would be observable given our observations.  We find that 59\% of companions to KOI 961 would have eluded detection.  Assuming an overall 40\% binary fraction for M dwarfs \citep[e.g.][]{Fischer1992,Covey2008}, this yields a probability of 0.24 that KOI 961 has a undetected companion.

\subsection{False Positive Probability}

\begin{table*}
\begin{center}
\caption{False Positive Probabilities for KOI 961}
\begin{tabular}{c c c c c} 
\hline\hline                        
 & BGEB\footnotemark[1] &  HEB\footnotemark[2] &  BGpl\footnotemark[3] & FPP\footnotemark[4]\\
\hline
961.01 & 9.7e-05 & 3.8e-05 & 5.4e-04 & 6.7e-04\\
961.02 & 2.8e-04 & 1.5e-05 & 1.1e-03 & 1.4e-03\\
961.03 & 1.1e-03 & 1.2e-03 & 1.1e-03 & 3.4e-03\\
\hline
\footnotetext[1]{Background eclipsing binary}
\footnotetext[2]{Hierarchical eclipsing binary}
\footnotetext[3]{Transiting planet around a background star}
\footnotetext[4]{Total false positive probability}
\label{FPPtable}
\end{tabular}
\end{center}
\end{table*}

\begin{figure*}[htbp]
   \centering
   \includegraphics[width=7in]{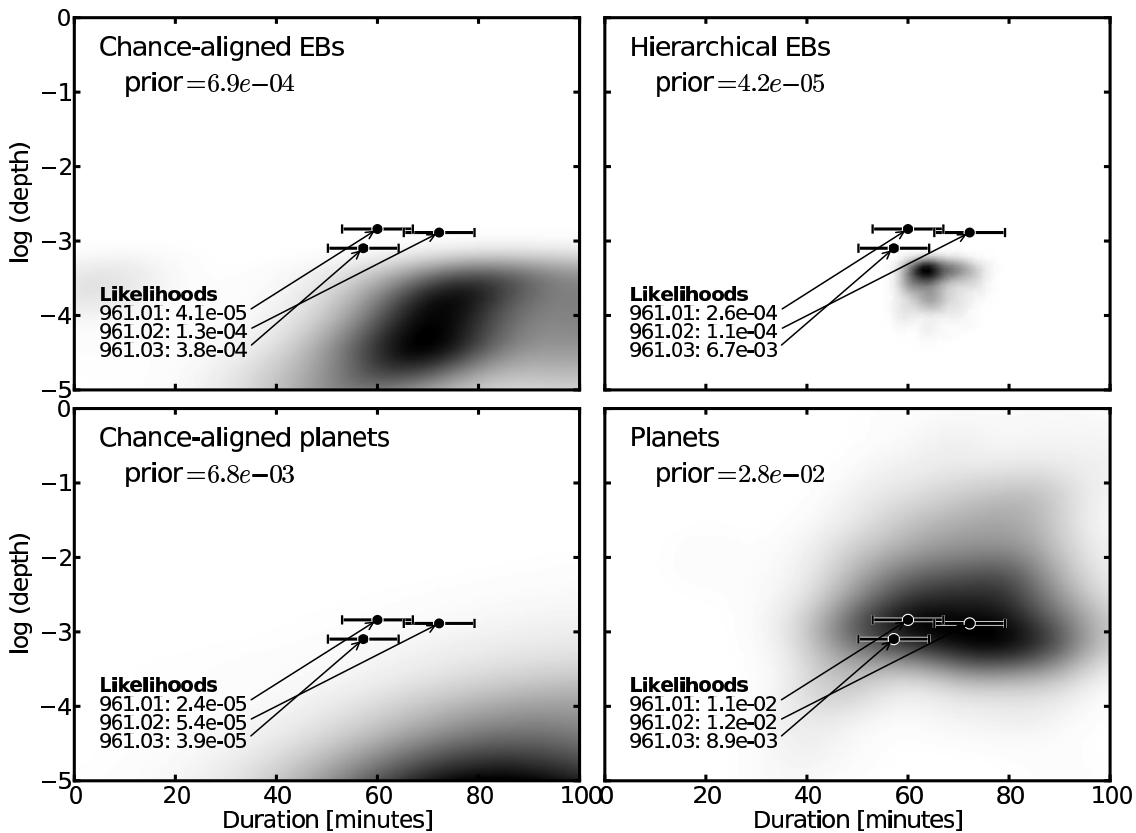}
   \caption{The probability density as a function of observed transit duration and depth for realistic populations of the three astrophysical false positive scenarios discussed in the text as well as the {\it bona fide} planet scenario.  We choose a statistically representative sample of configurations and model the light curves using \citet{MandelAgol2002}, convolved with a 29.4 minute cadence, assuming all circular orbits.  Included in these density plots are only those scenarios in which secondary eclipses would not be detectable in the KOI 961 light curve.  The observed transit signals of KOIs 961.01, 961.02, and 961.03 are marked (with durations scaled to match the period of 961.01, which was the basis for the population simulations), and do not fall within a high-likelihood region for any of the false positive scenarios.  The relative probability for each planet to be caused by each scenario is the product of the prior and likelihood for that scenario.}
   \label{fig:FPfig}
\end{figure*}

Calculating the false positive probability (FPP) for a particular transit signal requires comparing the probabilities of astrophysical false positive scenarios with an assumed probability of the planet hypothesis.  To this end, we calculate the probability that each of the potentially blending scenarios discussed above might be an eclipsing system mimicking a planet-like transit signal around KOI 961.  

Direct application of the exact results of \citet{Morton2011} is not appropriate for these signals, because the transit shape does not immediately rule out grazing eclipses.  As we show in Section \ref{transits}, the 29.4-minute observing cadence of the public {\it Kepler} data strongly affects the apparent shape of the light curve, including its depth (see Figure \ref{transits}).  Instead, we extend the methods of \citet{Morton2011} tailored to the specifics of the KOI 961 system.

The FPP for any of the three signals may then be written as follows:
\begin{equation}
\label{eq:FPP}
{\rm FPP} = 1 - \frac{\pi_{\rm pl} \mathcal L_{\rm pl}(\delta,T)}{\displaystyle \sum_i \pi_i \mathcal L_i(\delta,T)},
\end{equation}
where the index $i$ represents each of the false positive scenarios and the planet hypothesis, $\mathcal L_i(\delta,T)$ is the probability density function (PDF) for the given scenario evaluated at the specific depth and duration of the observed signal (the ``likelihood''), and $\pi_i$ is the {\it a priori} probability of each scenario existing.

The {\it a priori} probability for each scenario has two components: first, that the particular blending star configuration exists; second, that it eclipses and can mimic a planet around KOI 961.  For the background eclipsing binary (BGEB) scenario, we assume that a chance-aligned star has a 50\% chance of being a binary, and a 10\% chance of that binary being on a short-period orbit \citep{Raghavan2010}, using $<$ 50 days as a definition of ``short-period'' to allow convenient comparison to planet occurrence studies.  Using 0.18 as a conservative average eclipse probability (corresponding to the grazing eclipse probability of two 0.5 $R_\odot$ stars orbiting each other with a 1.2 day period), this gives a probability of $0.13 \times 0.50 \times 0.1 \times 0.18 = 1.1 \times 10^{-3}$ that there is a BGEB within the {\it Kepler} apertures of KOI 961.  In addition, the binary must not show a secondary eclipse in the diluted light curve.  We determine by inspection that a signal with a depth of $5\times 10^{-4}$ would have been detectable by Kepler, and so we use that as a conservative limit for the maximum depth of any putative diluted secondary eclipse.  We find that this disqualifies 39\% of BGEBs, resulting in $\pi_{\rm BGEB} = 6.9 \times 10^{-4}$.   

To estimate the prior probability of a hierarchical triple eclipsing binary (HEB) system, we assume that a potential companion to KOI 961 has a 40\% chance of having a companion of its own, a 20\% chance of that companion being on a short orbit [we conservatively use a higher short-orbit probability for a hierarchical system than for an isolated binary motivated by the observation that many close binaries have distant companions \citep{Tokovinin2006}], and an eclipse probability of 0.10 (corresponding to two 0.15 $\rm R_\odot$ stars in a 1.2 day orbit).  Furthermore, we simulate a population of HEBs according to a flat mass ratio distribution and find that only 2.3\% of HEB scenarios have diluted secondary eclipses shallower than $5 \times 10^{-4}$. This results in $\pi_{\rm HEB} = 0.24 \times 0.40 \times 0.20 \times 0.10 \times 0.023 = 4.4 \times 10^{-5}$.

For the background transiting planet hypothesis (BGpl),  we assume an overall 40\% occurrence rate for planets with periods $<$ 50 days \citep{Howard2010,Howard2011,Bonfils2011}, and an average transit probability of 0.13 (corresponding to a planet orbiting a 0.5 $R_\odot$ star in a 1.2 day period), giving $\pi_{\rm BGpl} = 6.8\times 10^{-3}$.  These same assumptions (except for a transit probability of 0.07 corresponding to the radius of KOI 961) yield an {\it a priori} probability for the {\it bona fide} KOI 961 transiting planet hypothesis of $\pi_{\rm pl} = 0.40 \times 0.07 = 0.028$.  We note that we have assumed isotropically oriented orbits for all the eclipse and transit probabilities we have used. This may not necessarily be the case for multiple planet systems (orbits may be coplanar) or for hierarchical systems (orbits of a hierarchical binary may be aligned with the plane of the planets), but assuming isotropic orientation for all orbital planes is a conservative approach.

To calculate the likelihood factors for each of the scenarios, we simulate the expected population of light curves following \citet{MandelAgol2002}, allowing for all impact parameters (including grazing eclipses), including limb darkening, and accounting for the effects of the 29.4-minute observing cadence.  In contrast to the BLENDER procedure \citep{Torres2011,Fressin2011,Fressin2011b}, we do not fit false positive light curves to the data; instead we choose the scenarios to simulate according to their {\it a priori} probability of existence, which we determine using the methodology of \citet{Morton2011}.  This allows us to directly extract the joint PDF of duration and depth for each scenario, taking care to include only those light curves with undetectable diluted secondary eclipses.   We also create the same distribution for the planet scenarios, assuming a population of planets with a radius distribution $dN/dR_p \sim R_p^{-2}$ from 0.4 to 20 $R_\oplus$, and accounting for the mass and radius errors of KOI 961.  The result is exactly analogous to Figure 5 of \citet{Morton2011} extended into the duration dimension.  ``Duration'' and ``depth'' for our purposes here are determined by inspection, not by any fitting procedure; ``duration'' is the total time the transit model flux spends below unity, and ``depth'' is simply the minimum value of the model light curve during primary transit.  These values correspond to what one would naively call duration and depth from inspecting a real light curve by eye; they have no direct physical significance except for providing a convenient, model-independent way to parameterize a light curve.  We then evaluate the PDFs at the duration and depth of each of the three candidate signals (integrating over errors in both dimensions) to obtain each of the likelihood factors (i.e.~the ``probability of the data given the model'').  The simulations we use to create the PDFs are all based on the period of KOI 961.01, so in order to evaluate the likelihoods for 961.02 and 961.03 at the proper locations we scale their true durations by a factor of $(P/1.21{\rm d})^{1/3}$.  Figure \ref{fig:FPfig} illustrates each of these PDFs and shows the likelihood value for each candidate for each scenario.

Table \ref{FPPtable} lists the normalized probabilities ($\pi_i \mathcal L_i/\sum \pi_i \mathcal L_i$) for each of the false positive scenarios for each of the candidates as well as the overall FPP.  The total false positive probabilities for 961.01, 961.02, and 961.03 are $6.7 \times 10^{-4}$ ($\sim$1 in 1500), $1.4\times 10^{-3}$ ($\sim$1 in 700), and $3.4 \times 10^{-3}$ ($\sim$1 in 290), respectively.  We note that these calculations have used conservative assumptions at every stage.  In particular we have ignored the fact that the ensemble evidence of the {\it Kepler} planet sample points to coplanarity being common among close-in, tightly packed planetary systems \citep{Lissauer2011}; if we properly accounted for this effect, the planet prior for each of the candidates would increase (corresponding to an increased transit probability), leading to correspondingly lower FPPs.  We therefore consider all three candidates in the KOI 961 system to be statistically validated planets, according to the standards set in the literature by the {\it Kepler} team.

\subsection{Misclassified planet probability}
We note that even though we have demonstrated all three transit signals in the KOI 961 system to be of planetary origin, there is still some probability of their radii being underestimated due to the presence of diluting flux within the aperture.  Specifically, there are three sub-scenarios that are encompassed by the ``planet in the KOI 961 system'' scenario:

\begin{enumerate}
\renewcommand{\labelenumi}{(\alph{enumi})}
\item KOI 961.0x orbits KOI 961, which is a single star.
\item KOI 961.0x orbits the primary component of KOI 961, which is an unresolved binary star system.
\item KOI 961.0x orbits the secondary component of KOI 961, which is an unresolved binary star system.
\end{enumerate}

\begin{figure*}[htbp]
   \centering
   \includegraphics[width=5in]{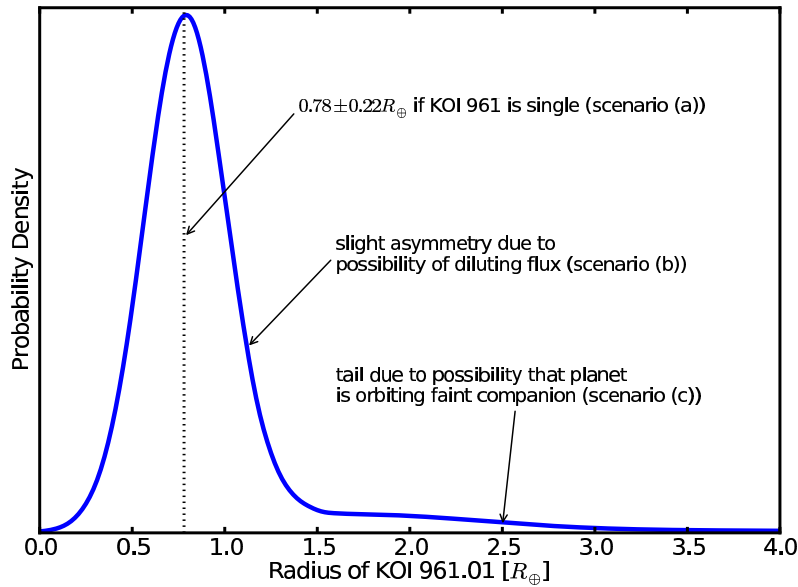}
   \caption{The probability distribution for the radius of KOI 961.01, accounting for potential binary star configurations of KOI 961 that could introduce diluting flux within the {\it Kepler} aperture.  Scenario (a) is that KOI 961 is a single star (probability 0.76), scenario (b) is that KOI 961 is a binary with the planet transiting the brighter component (probability 0.12), and scenario (c) is that KOI 961 is a binary with the planet transiting the fainter component (probability 0.12).  The planets 961.02 and 961.03 have analogous probability distributions.}
   \label{fig:rpdf}
\end{figure*}

We calculated in the previous subsection that there was a 24\% chance of KOI 961 having an undetected binary companion, giving scenario (a) a probability of 0.76.  We then assume that if KOI 961 is a binary, there is equal probability for a planet to be orbiting either component, giving both (b) and (c) probabilities of 0.12.  Using the same binary companion simulations discussed above, we calculate the effect this has on the radii of the planets, according to the following:
\begin{equation}
\label{eq:dilutedr}
R_{\rm p,true} = R_{\rm p, inferred} \left( 1 + 10^{-0.4 \Delta M_{\rm Kep}}\right) ^ {\frac{1}{2}},
\end{equation}
where $R_{\rm p,true}$ is the true planet radius, $R_{\rm p, inferred}$ is the radius inferred assuming the star is single, and  $\Delta M_{\rm Kep}$ is the magnitude difference between the two components (which would be negative if the planet is orbiting the fainter star).  As each binary companion in our simulation has an assigned {\it Kepler} magnitude, we can turn this putative population into a probability distribution function for the true planet radius, which we plot in Figure \ref{fig:rpdf} for KOI 961.01.  73\% of the probability mass remains below 1 $R_\oplus$, compared to 84\% of probability mass for the (a) scenario probability distribution alone.  The planets 961.02 and 961.03 have analogous probability distributions, with lower mean values, so we conclude that the possibility of a binary companion does not alter our conclusions of the sub-Earth-sized nature of the KOI 961 planets.  The resulting uncertainties in the planet radii due to mischaracterization are similar to the quoted uncertainties in Table \ref{planet_table}, which were calculated assuming scenario (a) and dominated by errors in the stellar parameters.  Therefore, we proceed assuming the uncertainties in Table \ref{planet_table}.

\section{Masses, Dynamics and Architecture}\label{dynamics}

\begin{figure}[]
\begin{center}
\includegraphics[width=3.3in]{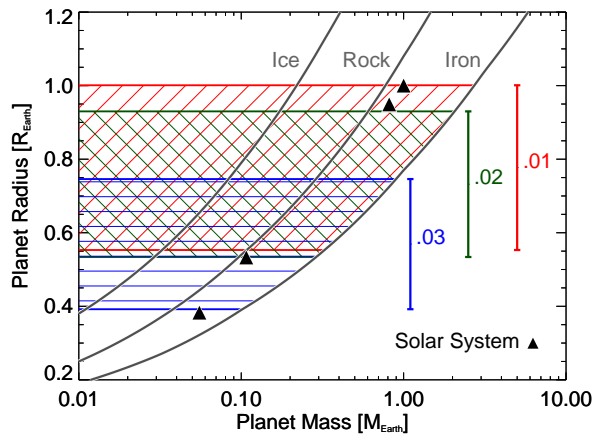}
\caption{Planetary Mass vs. Radius showing the $\pm 1\sigma$ values for the KOI 961 planetary radii and the range of possible masses.  We show theoretical models for cold spheres composed of pure iron, rock ($\rm Mg_2SiO_4$) and water-ice, interpolated from Table 1 of \citet{Fortney2007}.  We also include Solar System bodies Mercury, Mars, Venus and Earth, from left to right ({\it black triangles}).  Planetary bodies are unlikely to contain significant amounts of elements that are heavier than iron, so we use the pure iron mass-radius contour to place upper limits on the masses KOI 961.01 ({\it red}), .02 ({\it green}) and 0.3 ({\it blue}).}
\label{mass_radius}
\end{center}
\end{figure}

We do not currently have measurements of the planets' masses, as KOI 961 is faint ($V$ = 16.1 $\pm$ 0.055) and the expected radial-velocity semi-amplitudes are small: $K$ = 1.1, 1.25 and 0.33 m $\rm s^{-1}$ for KOI 961.01, 961.02 and 961.03 respectively.  In Figure \ref{mass_radius} we show a planetary mass-radius diagram with theoretical predictions for cold spheres composed of pure iron, rock ($\rm Mg_2SiO_4$) and  water-ice, interpolated from Table 1 of \citet{Fortney2007}, and we show potential masses for KOI 961's three planets.  Planetary bodies are unlikely to be dominated by elements that are heavier than iron; therefore, the pure iron mass-radius relation provides firm upper limits to the planet masses.  We report 1$\sigma$ upper limits of 2.73, 2.06 and 0.90 $\rm M_\Earth$ for KOI 961.01, .02 and .03 respectively.  With such low-masses, we believe it is unlikely that the planets could retain gaseous atmospheres with scale-heights that are a significant fraction of the planets' opaque radii.  The range of possible gaseous scale-heights for a specified radius, surface gravity and equilibrium temperature provide a lower-limit on the planets masses; however, we leave this calculation for a future paper.

To consider the dynamical effects in this multiple-planet system, we assume a stellar mass 0.14 $M_\sun$, and the 1-$\sigma$ upper limits on planetary masses.  We considered coplanar orbits, as this is highly likely due to the multi-transiting nature of this system (Lissauer et al. 2011).  In the following calculations we also assume near circular orbits.  The eccentricity damping timescale due to tides raised on the planets due to the star is \citep{2002Wu} :

\begin{equation}
\tau_e = \frac{2}{63 \pi}  P Q' \frac{M_p}{M_\star} ( \frac{a}{R_\star} )^5, 
\end{equation}

where Q' is a parameter that is $\lesssim 10^3$ for terrestrial bodies \citep{2004Mardling}.  Using measured values for $P$, $a$, $M_\star$, and $R_\star$, and assuming a terrestrial density to estimate $M_p$, all the planets have short eccentricity-damping timescales compared to the age of the system ($>$ 4.5 Gyr based on H-$\alpha$), even the outermost planet ($\tau_e \simeq$ 0.1 Gyr).  This justifies the assumption of circular orbits in modeling the present-day dynamics of the system. 

Given the very low-mass planets we have inferred, and the distances from strong mean-motion resonances, we would not expect large transit timing signatures.  We performed 3-body numerical integrations to determine simulated transit times using the method of \citet{2010Fabrycky}.  Each simulation had two planets -- their perturbations were considered pairwise.  The strongest interaction was between KOI-961.01 and KOI-961.03, with peak-to-peak deviations from a linear ephemeris of 0.7 minutes and 2.6 minutes, respectively.  Other pairs showed less than 1 second of deviation.  Although KOI-961.03's deviation may be detected in the full Kepler data set, we expect other interactions to be undetectable.

We also investigated the long-term stability of this system.  First, we compute the separation in number of Hill spheres \citep{1996Chambers} between the inner two planets and between the outer two planets: they are 19.5 and 9.6, respectively.  According to \citet{1993Gladman}, pairs of planets are Hill stable (i.e. orbits that are initially circular may not cross) if this value is greater than 3.46.  When three-planet systems are considered, the stability boundary moves to larger separation; for instance, \citet{1996Chambers}[fig. 3] shows instability is possible in $\sim10^6$~orbits if the Hill separation between pairs of planets in three-planet systems is 6.  For our case, scaling relations suggest that stability should easily be achieved in this three-planet system \citep{1996Chambers}.  Moreover, \citet{Lissauer2011} integrated three-planet systems with precession mimicking the general relativistic effects, including a model for KOI-961 with much larger mass ratios than we suggest.  They found stability for over 10 billion orbits. Future work could consider precession due to tidal and rotational effects and initially non-circular orbits.

\section{Summary and Discussion}\label{discussion}

The {\it Kepler} Mission has demonstrated unprecedented photometric precision over long time baselines, opening up opportunities to study planets with radii comparable to the Solar System terrestrial planets \citep{Borucki2011,Howard2011}. However, a transit light curve does not provide a direct measure of a planetÕs radius, but instead gives only the ratio of planet and stellar radii. Thus, the precision of the {\it Kepler} transit light curves cannot be fully exploited without precise and accurate knowledge of the stellar radii of the host stars.

This problem is particularly acute for the planetary system around KOI-961. Because the host star is an M dwarf, the photometry-based stellar properties in the {\it Kepler} Input Catalog suffer from large systematic errors. However, we have taken advantage of two opportunities afforded by the nature of this star. First, we find that KOI-961 is a near twin to a nearby, well-studied M-dwarf.  By performing a differential analysis with respect to BarnardÕs star, we find that KOI-961 is also a small star with large proper motion.  Second, the high proper motion greatly facilitates the validation of the planetary system based on examination of historic imaging data. We find that there are no potential eclipsing binaries at the current position of KOI-961 up to 4.8 magnitudes fainter.

Using revised and refined stellar parameter for KOI-961 we perform a reanalysis of the transit light curves, confirming the three signals at the periods reported by \citet{Borucki2011}. We revise the impact parameters and transit depths for all three planets. The transit depths combined with our estimate of the stellar radius yield planet radii that are all less than the radius of the Earth.  We measure radii of 0.78 $\pm$ 0.22 $\rm R_\Earth$ for KOI 961.01, 0.73 $\pm$ 0.20 $\rm R_\Earth$ for KOI 961.02, and 0.57 $\pm$ 0.18 $\rm R_\Earth$ for KOI 961.03.  

Incorporating all available observational constraints on the presence of potential blends (both chance-alignment and hierarchical), we are able to establish that the transit signals are indeed planets in the KOI 961 system.  We also quantify the probability that KOI 961 might be a binary with planets transiting either of the two components (which would lead to an underestimate of the planet radii), and find that this probability is low enough to remain confident that we have not drastically misclassified the planets.  Follow-up adaptive optics imaging of KOI 961 will be able to put even tighter constraints on this possibility.

As a planet-hosting star, KOI 961 provides an important data point for the study of the planet-metallicity relationship in the context of low-mass stars and terrestrial-sized planets. The planet-metallicity correlation was originally discovered for Sun-like F, G and K-type stars, and correlated the stellar metallicity to the likelihood of hosting a Jovian-mass planet \citep[e.g.][]{Santos2001, Fischer2005, Johnson2010}.   Recently developed methods for estimating the metallicities of M dwarfs have shown that M-dwarf planet-hosts are also systematically metal-rich \citep{Johnson2009,Rojas2010,Schlaufman2010,Neves2011}.  Being significantly metal-poor, KOI 961 represents an outlier in the context of these studies, but it is not clear that the planet-metallicity correlation extends to stars which host Earth-mass and smaller exoplanets.  Recent analysis by \citet{Mayor2011} of radial-velocity detected exoplanets indicates that the correlation may not apply to stars which only host planets less-massive than Neptune.  However, there is evidence that the planet-metallicity correlation is important for small-planets around low-mass stars, based on a comparison of the KOIs to {\it Kepler} field stars by \citet{Schlaufman2011}.  The discovery of KOI 961, and the studies of \citet{Mayor2011} and \citet{Schlaufman2011}, strongly motivate further investigations of the planet-metallicity correlation for low-mass planets around low-mass stars.

\begin{figure}[]
\begin{center}
\includegraphics[width=3.3in]{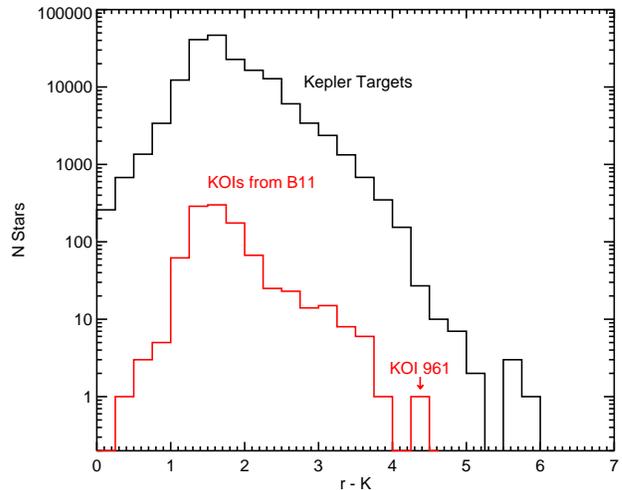}
\caption{Distribution of $r$-$K$ for the {\it Kepler} targets from the {\it Kepler} Input Catalog \citep[{\it black}, ][]{Batalha2010, Brown2011} and for the KOIs \citep[{\it red}, ][]{Borucki2011}, with a bin width of 0.25.  We have removed all objects with $J$ - $K$ $>$ $1.0$, as these are likely distant giant stars with significantly reddened colors.  KOI 961 is one of only a few dozen mid-to-late M dwarfs being observed by {\it Kepler}.}
\label{r_k}
\end{center}
\end{figure}

The detection of the planetary system around KOI 961 is remarkable since it is one of only a few dozen mid-to-late M dwarfs currently observed by {\it Kepler}.  Figure \ref{r_k} shows the $r$-$K$ distribution for all targets currently observed by {\it Kepler}, and KOIs in \citet{Borucki2011}.  We have removed all objects with $J$ - $K$ $>$ $1.0$, as these are likely distant giant stars with significantly reddened colors (see Figure \ref{j_k}).  KOI 961 is the reddest dwarf KOI from the \citet{Borucki2011} release, and is also likely the least massive KOI.  Only 27 dwarf stars observed by {\it Kepler} have $r$-$K$ greater than KOI 961's value of 4.444.  Combined with the low probability of a planetary system being geometrically aligned such that a transit is observed (only 13\% in the case of KOI 961.01), {\it Kepler}'s discovery of planets around KOI 961 could be an indication that planets are {\it common} around mid-to-late M dwarfs, or at least not rare.  This would be consistent with the results of \citet{Howard2011}, who used the {\it Kepler} detections to show that the frequency of sub-Neptune-size planets ($R_P$ = 2-4 $\rm R_\Earth$) increases with decreasing stellar effective temperature for stars earlier than M0.  Results from on-going exoplanet surveys of M dwarfs such as MEarth \citep[e.g.][]{Irwin2011}, and future programs such as the Habitable Zone Planet Finder \citep[e.g.][]{Mahadevan2010, Ramsey2008} and CARMENES \citep{Quirrenbach2010}, will shed light on the statistics of low-mass planets around mid-and later M dwarfs.

\acknowledgements

We would like to thank Peter Dawson, who provided the multi-wavelength spectrum of Barnard's Star.  We would like to thank Sarah Ballard and Michael Line for the thoughtful discussions concerning the paper.  We would like to thank Bruce Gary of Hereford Arizona Observatory for taking the $B$,$V$ and $R_C$ photometric measurements of KOI 961.

This work includes observations taken at the Palomar Obervatory 200-inch Hale Telescope granted by Cornell University.  The TripleSpec spectrograph was built at Cornell and delivered to Palomar as part of the Cornell-Caltech-Palomar arrangement.  This work includes observations taken at the W. M. Keck Observatory and Palomar Observatory granted by the California Institute of Technology.  Some of the Palomar 200-inch Telescope time was provided by NASA/JPL.  We thank Melodie Kao, Matthew Giguere and Ming Zhao for assisting with the Keck observations.

J.A.C, K.R.C., D.C.F. and E.N.K. acknowledge support for this work from the Hubble Fellowship Program, provided by NASA through Hubble Fellowship grants HF-51267.01-A, HST-HF-51253.01-A, HF-51272.01-A  and HST-HF-51256.01-A, awarded by the Space Telescope Science Institute, which is operated by the AURA, Inc., for NASA, under contract NAS 5-26555.  K.G.S., J.P. and L.H. acknowledge support through the Vanderbilt Initiative in Data-intensive Astrophysics and NSF grants AST-0849736 and AST-1009810.

This paper includes data collected by the Kepler mission. Funding for the Kepler mission is provided by the NASA Science Mission directorate.  Some of the data presented in this paper were obtained from the Multimission Archive at the Space Telescope Science Institute (MAST). STScI is operated by the Association of Universities for Research in Astronomy, Inc., under NASA contract NAS5-26555. Support for MAST for non-HST data is provided by the NASA Office of Space Science via grant NNX09AF08G and by other grants and contracts.

The United Kingdom Infrared Telescope (UKIRT) is operated by the Joint Astronomy Centre on behalf of the Science and Technology Facilities Council of the U.K.

The Digitized Sky Surveys were produced at the Space Telescope Science Institute under U.S. Government grant NAG W-2166. The images of these surveys are based on photographic data obtained using the Oschin Schmidt Telescope on Palomar Mountain and the UK Schmidt Telescope. The plates were processed into the present compressed digital form with the permission of these institutions.

The National Geographic Society - Palomar Observatory Sky Atlas (POSS-I) was made by the California Institute of Technology with grants from the National Geographic Society.  The Second Palomar Observatory Sky Survey (POSS-II) was made by the California Institute of Technology with funds from the National Science Foundation, the National Geographic Society, the Sloan Foundation, the Samuel Oschin Foundation, and the Eastman Kodak Corporation.  The Oschin Schmidt Telescope is operated by the California Institute of Technology and Palomar Observatory.

\clearpage
\bibliographystyle{apj}
\bibliography{koi961}

\end{document}